# A Human-Checkable Four-Color Theorem Proof

André Luiz Barbosa
http://www.andrebarbosa.eti.br
Non-commercial projects: SimuPLC – PLC Simulator & LCE – Electric Commands Language

***Abstract***. *This paper presents a short and simple proof of the Four-Color Theorem that can be utterly checkable by human mathematicians, without computer assistance. The new key idea that has allowed it and the global structure of the proof are presented in the Introduction.*

## 1. Introduction

The Four-Color Theorem (4CT) [1, 16, 17] is a very beautiful discovery, embraces very deep math with innocent appearance, and furthermore has an epic and passionate history.

Circa a century and two decades since it was conjectured, finally its computer program-based proof was amazing (and even a technological feat in the 1970s [13, 14]). However, as this proof – even though it was highly developed later [2, 15] – is yet so very long and composed of separated pieces through excruciating details, it is still no amenable to complete human verification. [2]

This little paper presents a very shorter and simpler proof susceptible to fully human verification and total understanding [12] (an overview of the proof is presented in Section 3).

A reviewer has asked me: "– What is the new key idea that allows us to get around looking into all these many different configurations that the existing proofs test for?"

The new key idea here is to see a map not as formed of possibly *so many different*

*configurations*, but simply see it as formed of *essentially only two closed curves* overlapping in a plane: then, all the regions from the map can only be in exactly a single position with respect to these two curves, from four possible options: out of them, inside of them, out of one and inside the other, or out of the other and inside the one, which naturally generates the four colors that they can be colored, without adjacent regions having the same color, since crossing a boundary from the map must necessarily change one, or the two, from the out/inside positions. Consequently, adjacent regions cannot have the same position – equivalently, the same color – with respect to the two original [set of] closed curves.

Hence, as only four different positions are possible with respect to these two (set of) closed curves (out-out, out-in, in-out, in-in), if we prove that any arbitrary hypothetical minimal counter-example to 4CT can always be formed by means of this overlapping construction (hence, by contradiction, that *hypothetical* minimal counter-example cannot exist as a *real* counter-example), then the 4CT stands proved, by *Reductio ad Absurdum*.

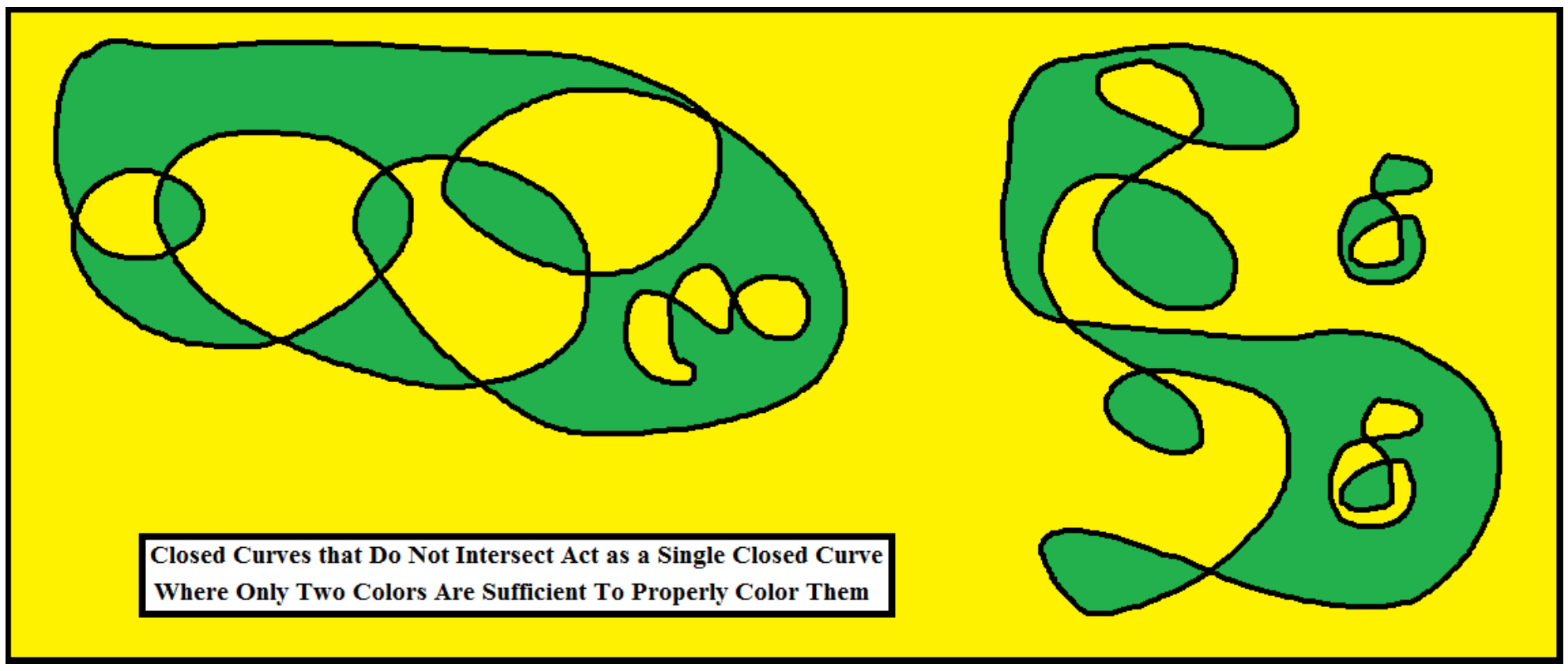

**Figure 1.1** Example 1 of Closed Curves Representing the New Key Idea Used in the Proof

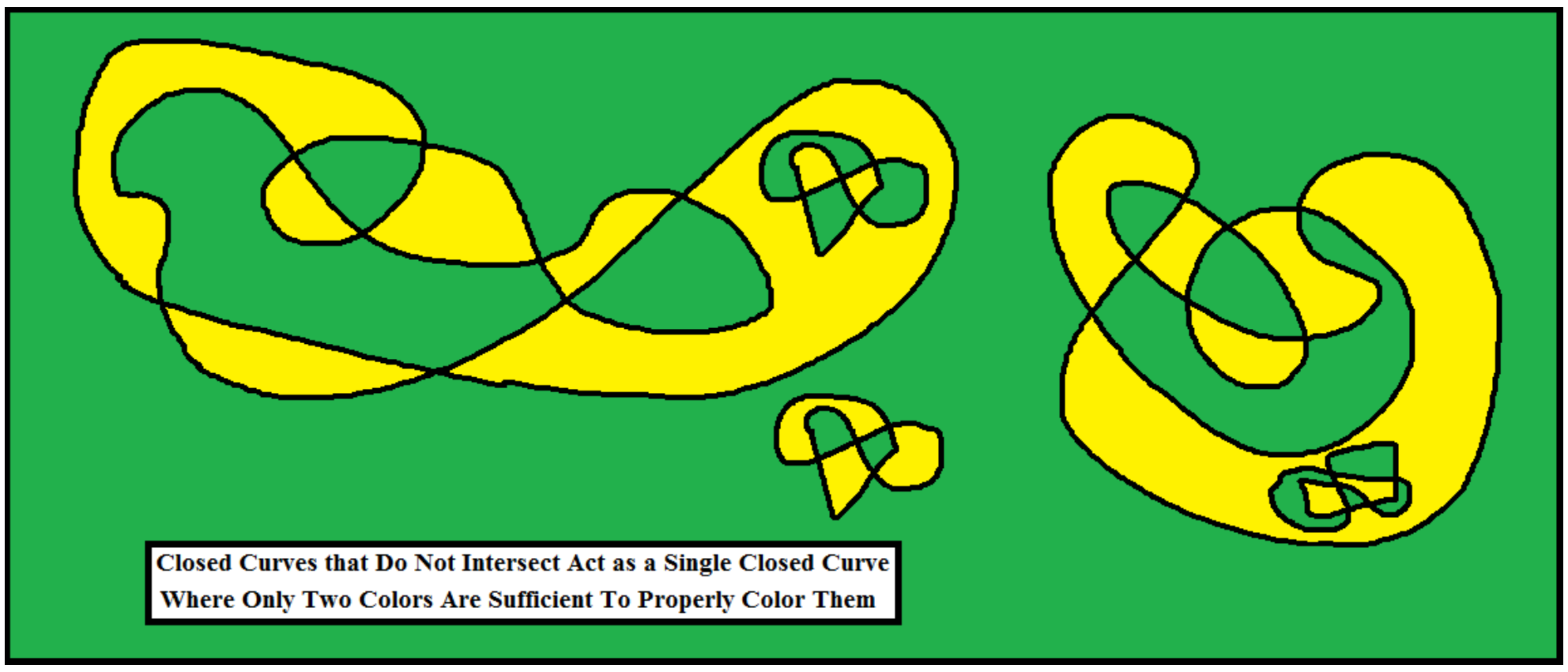

**Figure 1.2** Example 2 of Closed Curves Representing the New Key Idea Used in the Proof

With respect to two figures above, we can intuitively see that a resultant map from overlapping two [no intersecting or disjoint set of] closed curves can always be properly colored with only four different colors: All its regions can be colored with one from the four ones: green-green (1), green-yellow (2), yellow-green (3) or yellow-yellow (4) according to respective colors of original regions from two previous (set of) closed curves.

## 1.1 The Global Structure of the Proof

The global structure of the proof is constituted of the following six sequential steps:

i) Defining *2-DSCC_M*, *4-CM*, *3-ECC*, *CBG* and related objects.

ii) Proving that *2-DSCC_M* is equivalent to *4-CM*: *2-DSCC_M* ⇔ *4-CM*.

iii) Defining *3-ECC* and noting that every *3-ECC* is a *4-CM*, and a *2-DSCC_M* too: (a known result, really: *3-ECC* ⇒ *4-CM*, and so *3-ECC* ⇒ *2-DSCC_M*).

iv) Observing that an arbitrary hypothetical *minimal CBG* that is a counter-example to the 4CT would be a *CBG* ***N*** (an arbitrary *minimal* counter-example to the 4CT that is (or is converted into) a *CBG*) that has a pentagon inside it.

v) Proving that the resultant map ***C*** when that pentagon is removed from ***N*** is a *2-DSCC_M*, and a *3-ECC* too, beyond to be a *4-CM* (since ***N*** is minimal).

vi) Proving after all that we can, however, return that pentagon to map ***C*** generating newly that *CBG* ***N***, and then proving it is really a *3-ECC*, hence a *4-CM*, therefore proving that ***N*** was not really a minimal counter-example to 4CT, then such a minimal counter-example cannot exist, hence proving 4CT.

## 2. A Human-Checkable Four-Color Theorem Proof

**Definition 2.1. Disjoint Set of Closed Curves (*DSCC*).** A ***DSCC*** is a finite set of disjoint oriented closed curves [either simple (*Jordan curve*) or non-simple ones] in a Euclidean plane [3], where there is no intersection point between any pair of them – that is, they do neither cross, nor intersect, nor "*touch*" themselves in that set. The empty set and a unitary set of a simple closed curve are trivially *DSCC*.

More formally: Let $I_1, ..., I_n$ be nonempty intervals of reals $\mathbf{R}$, and $C_1$: $I_1 \rightarrow \mathbf{R}^2$, ..., $C_n$: $I_n \rightarrow \mathbf{R}^2$ be continuous mapping, where $I_i = [a_i, b_i]$, $b_i > a_i$, $(C_i(x) = C_i(y)) \wedge (x \neq y)$ only in finitely many $x$, $C_i(a_i) = C_i(b_i)$, $\forall i$, $1 \leq i \leq n$, $\{C_1, ..., C_n\}$ is a *DSCC* iff $C_i(x) = C_j(y) \Rightarrow i = j$, $\forall i, j$, $1 \leq i, j \leq n$.

**Definition 2.2. *DSCC* Winding Number (DWN).** The ***DWN*** of every point $(x, y)$ inside a Euclidean plane where the respective *DSCC* is drawn is the algebraic sum of all absolute values of the *winding numbers* (as defined in [4, 8]) of that point with respect to all curves $C_i$ from that *DSCC* (so, the orientations of the curves of the *DSCC* do not matter at all). Note that every arbitrary point of a Euclidean plane has a determined, fixed and effectively calculable winding number with respect to a determined oriented closed curve, so a DWN too, with respect to a given *DSCC*: DWN$(x, y) = \sum |$winding number$(x, y, C_i)|$.

**Definition 2.3. Points inside and outside with respect to a *DSCC*.** Every point $(x, y)$ inside a Euclidean plane with respect to a *DSCC* (where it is in) can be exactly and exclusively in only one of the following position:

(i) *Outside from the DSCC*, iff its respective DWN is even [$2 \mid$ DWN$(x, y)$];
(ii) *Inside from the DSCC*, iff its respective DWN is odd [$2 \nmid$ DWN$(x, y)$]; or
(iii) Inside the image of some curve of the *DSCC* (when DWN$(x, y)$ is undefined).

**Definition 2.4. Regions inside and outside from a *DSCC*.** An arbitrary region $R$ (a contiguous portion of surface that does not contain any point from any curve) in a Euclidean plane where a *DSCC* is drawn can be exactly and exclusively in only one of the following position: (Obs.: Some few small parts of the proof are visual – using geometric intuition –, but their truth are so obvious that they do not need formal combinatorics demonstrations in these few cases.)

(i) *Outside from the DSCC*, iff some point $(x, y) \in R$ is outside from that *DSCC*; or
(ii) *Inside from the DSCC*, iff some point $(x, y) \in R$ is inside from that *DSCC*.

Note: All the points from a specified region must have the same DWN, since the winding number of a continuously moving point with respect to some closed curve changes only if that point crosses (intersects) that closed curve. [4, 8] Thus, as in a region there is no point from the curves, the DWN of all its points are equal each other. Furthermore, by Def. 2.2, every region into a Euclidean plane has a determined, unique, fixed and effectively calculable position (*inside* or *outside*) with respect to a given *DSCC*.

Verify that the definitions of *outside* and *inside* above can be swapped without altering the essence of the proof in this paper.

**Definition 2.5. 2-*DSCC* Map (2-*DSCC*_M).** A ***2-DSCC_M*** is a connected finite simple planar graph that can be represented (drawn) by two *DSCC*, where we can call them a *Blue DSCC* and a *Yellow DSCC* (which can be considered formed of *blue* and *yellow* closed curves, where curves of different colors can *touch* (have common points with) themselves, but the curves of same color cannot do it, by Def. 2.1).

Note yet that a *DSCC* in this definition can be the empty set, and, in a 2-*DSCC*_M, regions (countries), boundaries (borders, sides), and vertices (points where different boundaries *touch* themselves) can be represented by faces, edges (set E) and vertices (set V) in a finite simple planar graph $M = (V, E)$, respectively (where all the edges from $M$ can be represented by $e_{i\text{-}j} = \{v_i, v_j\}$, where $v_i, v_j \in V$ and $\{v_i, v_j\} \in E$).

**Definition 2.6. Blue, yellow and green edges (boundaries).** In a 2-*DSCC*_M represented by a finite simple planar graph $M = (V, E)$, every edge $\mathbf{e_{i\text{-}j}}$ representing a boundary formed only by a blue (respect., yellow) curve is a *blue* (respect., *yellow) edge*, and every edge $e_{k\text{-}l}$ representing boundary formed by an intersection of a blue and a yellow curve (overlapping infinitely many points) is a *green edge*. These occurrences can be represented by the predicates $B(e_{i\text{-}j})$ (respect., $Y(e_{i\text{-}j})$) and $G(e_{k\text{-}l})$, respectively.

In order to formally include these colors to the maps, we shall introduce two new planar graphs $B = (V, E_b)$, and $Y = (V, E_y)$, where a blue (respect., yellow) edge will be represented by $e_{bi\text{-}bj} = \{v_{bi}, v_{bj}\} \in E_b$ (respect., $e_{yk\text{-}yl} = \{v_{yk}, v_{yl}\} \in E_y$), and a green edge is defined as one that is in these two graphs at same time ($e_{bi\text{-}bj}$ and $e_{yi\text{-}yj}$), linking the same points $v_i$ and $v_j$ (where $v_{bi} = v_{yi} = v_i$, $\forall v_i \in V$, and $E_b \cup E_y = E$).

**Lemma 2.2.** Every edge of a ***2-DSCC_M*** is either a blue, yellow or green edge.

*Proof.* In a ***2-DSCC_M*** there are only blue and yellow curves, and every segment where they intersect is (must be) green (representing an overlap of the two curves). Hence, can there be only blue, yellow and green edges (boundaries) in a ***2-DSCC_M***. □

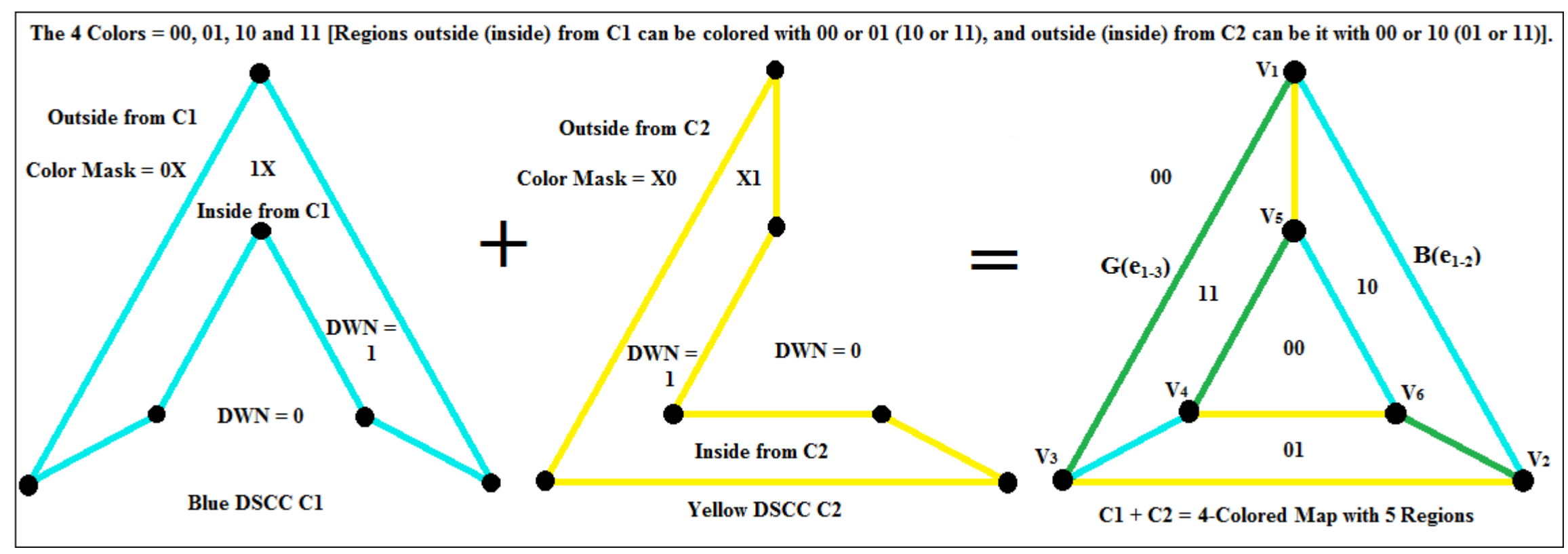

**Figure 2.1** Example of How 2 Closed Curves form a 4-Colored (or 3-Edge-Colored) 2-*DSCC*_M

More formally, the *2-DSCC_M* in the Fig. 2.1 can be represented by $M = (\{v_1, v_2, v_3, v_4, v_5, v_6\}, \{e_{b1\text{-}b2}, e_{b3\text{-}b4}, e_{b5\text{-}b6}, e_{y1\text{-}y5}, e_{y2\text{-}y3}, e_{y4\text{-}y6}, e_{g1\text{-}g3}, e_{g4\text{-}g5}, e_{g2\text{-}g6}\})$.

**Lemma 2.3.** All regions $R_i$ are exclusively either inside or outside from each one of the two *DSCC* of a ***2-DSCC_M*** (more formally, either $R_i$(*Blue DSCC*) = *Inside* or *Outside*, and either $R_i$(*Yellow DSCC*) = *Inside* or *Outside*).

*Proof.* All the points from a specified region must have the same DWN, since the winding number of a continuously moving point with respect to some closed curve changes only if that point crosses that closed curve. [4, 8] Since the intersection of a region with the curves from a ***2-DSCC_M*** is an empty set, this region is exclusively either inside or outside each curve from that ***2-DSCC_M***, by Def. 2.4:

$$(x_1, y_1), (x_2, y_2) \in R_i \Rightarrow \text{DWN}(x_1, y_1) = \text{DWN}(x_2, y_2). \ \square$$

**Definition 2.7. Four-colorable map (4-CM).** A ***4-CM*** is a *4-colorable* planar map (connected finite simple planar graph), that is, a planar map where at most four colors suffice to color it without adjacent regions having the same color. [1, 16, 17]

**Lemma 2.4.** Every ***2-DSCC_M*** is also a ***4-CM***.

*Proof.* We can color the regions in any *2-DSCC_M* as indicated in the table below:

| Is the Region Inside the *Blue DSCC*? | Is the Region Inside the *Yellow DSCC*? | Color the Region With the Color: |
|---|---|---|
| No | No | 00 (■) |
| No | Yes | 01 (■) |
| Yes | No | 10 (■) |
| Yes | Yes | 11 (■) |

**Table 2.1** How to properly 4-color a *2-DSCC_M* accordingly regions' positioning

Thus, within the Table 2.1, since there are in the *2-DSCC_M* two *DSCC*, and all the regions from the map are either inside or outside each curve, by Lemma 2.3, four colors are sufficient in order to color that map, in general, as though each region inside (respect., outside) the *Blue DSCC* had a color mask ***1X*** (respect., ***0X***), and each region inside (respect., outside) the *Yellow DSCC* had a color mask ***X1*** (respect., ***X0***), where the color of that region is like the composition of these two masks, as shown in the Table 2.1. But with only four colors, couldn't there be two adjacent regions with the same color? We shall see below that this flaw cannot occur:

If two adjacent regions had the same color, then the position of those regions would be the same one with respect to two *DSCC*, but this is impossible, since in order to cross the edges of the map we can only cross either exactly one blue or exactly one yellow curve, or exactly two ones at same time (when crossing a green edge), exactly one of each color, by Def. 2.1 (remember that the blue (respect., yellow) curves of a *DSCC* cannot have common edge, that is, infinitely many overlapped points – only finitely many [isolated] ones – with themselves); that is, when we intersect (cross) an edge, moving us from a region to another one, we must change at least one of the answers from Table 2.1 (from *No* to *Yes* or from *Yes* to *No*) with respect to the incoming region, by Def. 2.4; thus, it is impossible that two adjacent regions have the same set of those two positioning answers in that table above, so it is impossible those two adjacent regions have the same color.

Hence, the coloring of the map by means of the Table 2.1 guarantees that there is no adjacent region with the same color, so at most four color suffice to properly color every *2-DSCC_M*; thus, every *2-DSCC_M* is also a *4-CM*. □

**Lemma 2.5.** Every ***4-CM*** is also a ***2-DSCC_M***.

*Proof.* This proof is constructive, since we shall demonstrate that we can construct a *2-DSCC_M* from every *4-CM* properly 4-colored.

See, from any *4-CM* properly 4-colored *M*, we can decide which regions in it are either inside or outside each curve of a supposed *2-DSCC_M*, as indicated in the table below:

| Color of the Region In *M* | Is the Region Inside the *Blue DSCC*? | Is the Region Inside the *Yellow DSCC*? |
|---|---|---|
| 00 (🟥) | No | No |
| 01 (🟨) | No | Yes |
| 10 (🟦) | Yes | No |
| 11 (🟩) | Yes | Yes |

**Table 2.2** How to position regions in a *2-DSCC_M* accordingly their coloring

Thus, the *Blue DSCC* is formed by all edges adjacent to regions colored with color 10 (🟦) or 11 (🟩), when they are also adjacent to regions colored with some color from the other two ones, 00 (🟥) or 01 (🟨), and the *Yellow DSCC* is formed by all edges adjacent to regions colored with color 01 (🟨) or 11 (🟩), when they are also adjacent to regions colored with some color from the other two ones, 00 (🟥) or 10 (🟦).

Then, from the table above, we can decide which edges in every *4-CM* properly 4-colored form each curve of that supposed *2-DSCC_M*, as indicated in the table below:

| Edge Adjacent to Two Regions Colored With: | Is the Edge in the *Blue DSCC*? | Is the Edge in the *Yellow DSCC*? |
|---|---|---|
| 00 (🟥) \| 01 (🟨) | **No** | **Yes** |
| 00 (🟥) \| 10 (🟦) | **Yes** | **No** |
| 00 (🟥) \| 11 (🟩) | **Yes** | **Yes** |
| 01 (🟨) \| 10 (🟦) | **Yes** | **Yes** |
| 01 (🟨) \| 11 (🟩) | **Yes** | **No** |
| 10 (🟦) \| 11 (🟩) | **No** | **Yes** |

**Table 2.3** How to decide belonginess of edges w.r.t. the *DSCCs* accordingly their adjacency

**Proposition 2.1.** In order to construct the *Blue DSCC* of the map *M*, we can make a new map *M'* simply excluding from the original *4-CM* all the edges that are not in this *DSCC*, that is, those ones that are adjacent to two regions colored with 00 (🟥) and 01 (🟨), and those ones that are adjacent to two regions colored with 10 (🟦) and 11 (🟩), coloring the resultant regions of *M'* with 00 (🟥) or 10 (🟦), accordingly they encompass regions from *M* colored with 00 (🟥) and/or 01 (🟨), or with 10 (🟦) and/or 11 (🟩), respectively. Therefore, *M'* turns out to be a map properly *2-colored*. And all map (where the edges form closed curves, since there is no cut edges) of this type can be considered a *DSCC*, with the regions colored with a color representing regions outside of that *DSCC*, and the ones colored with the other color representing regions inside of the *DSCC*.

*Proof.* Excluding the edges adjacent to two regions colored with 00 (■) and 01 (■), or 10 (■) and 11 (■) from the map *M*, and maintaining the other ones, is equivalent to generate a map *M'* with only two types of regions: the 00-01 and the 10-11 ones, that we can color with (■) and (■), respectively, where *M'* is properly *2-colored*, since it is impossible that two regions of same color are adjacent in *M'*, for in order to this fact occur we would need maintain in *M'* at least one edge adjacent to two regions colored with 00 (■) and 01 (■), or 10 (■) and 11 (■), which is not permitted, by Proposition 2.1:

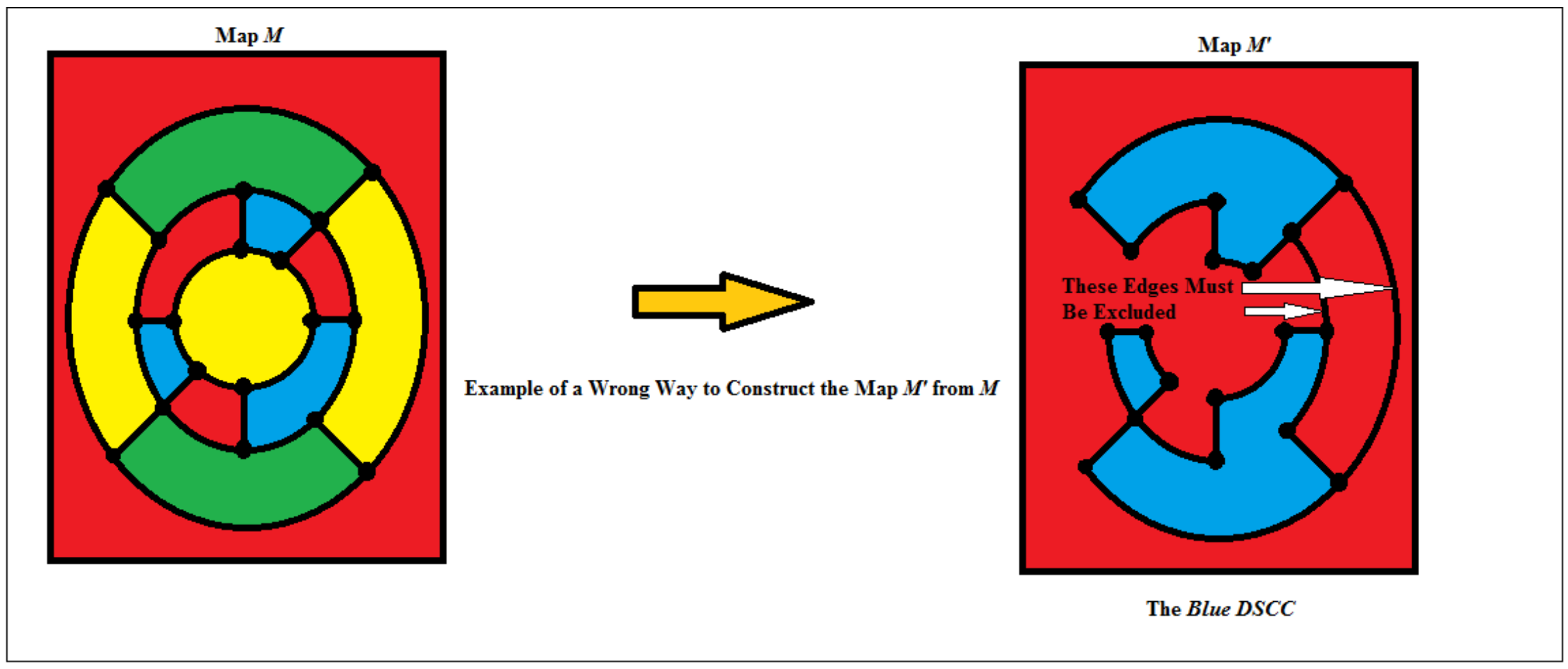

**Figure 2.2** Example of How *NOT* to Construct a *Blue and* a *Yellow DSCC* from a 4-CM

In order to illustrate the construction above, see that those remaining edges in *M'* form at least one closed curve, since all the vertices from a *2-colorable* map must have an even degree (otherwise, it could not be *2-colorable*), therefore they must form one or more *Eulerian cycles*, [5, 16, 17] which can be represented by a set of disjoint closed curves (since *Eulerian cycles* neither pass by any edge of the map more than once, nor cross other edges, so those closed curves do not cross other ones, obeying the Def. 2.1). Then, we can call this set of curves *the Blue DSCC* of the map. Similarly by symmetry, swapping the roles of the colors 10 (■) and 01 (■) in the above argument, we see that the same process shall build *the Yellow DSCC* of the *2-DSCC_M*.

Notice yet that all the edges of that *4-CM M* must be in some *DSCC*, since the conditions that the edges must obey when forming each one of the two *DSCCs* exhaust all the edges in the map (the edges excluded in the formation of a *DSCC* are not excluded in the formation of the other one, and vice versa, thus all the edges of the map shall eventually participate (must do it) in the formation of some *DSCC*, after all):

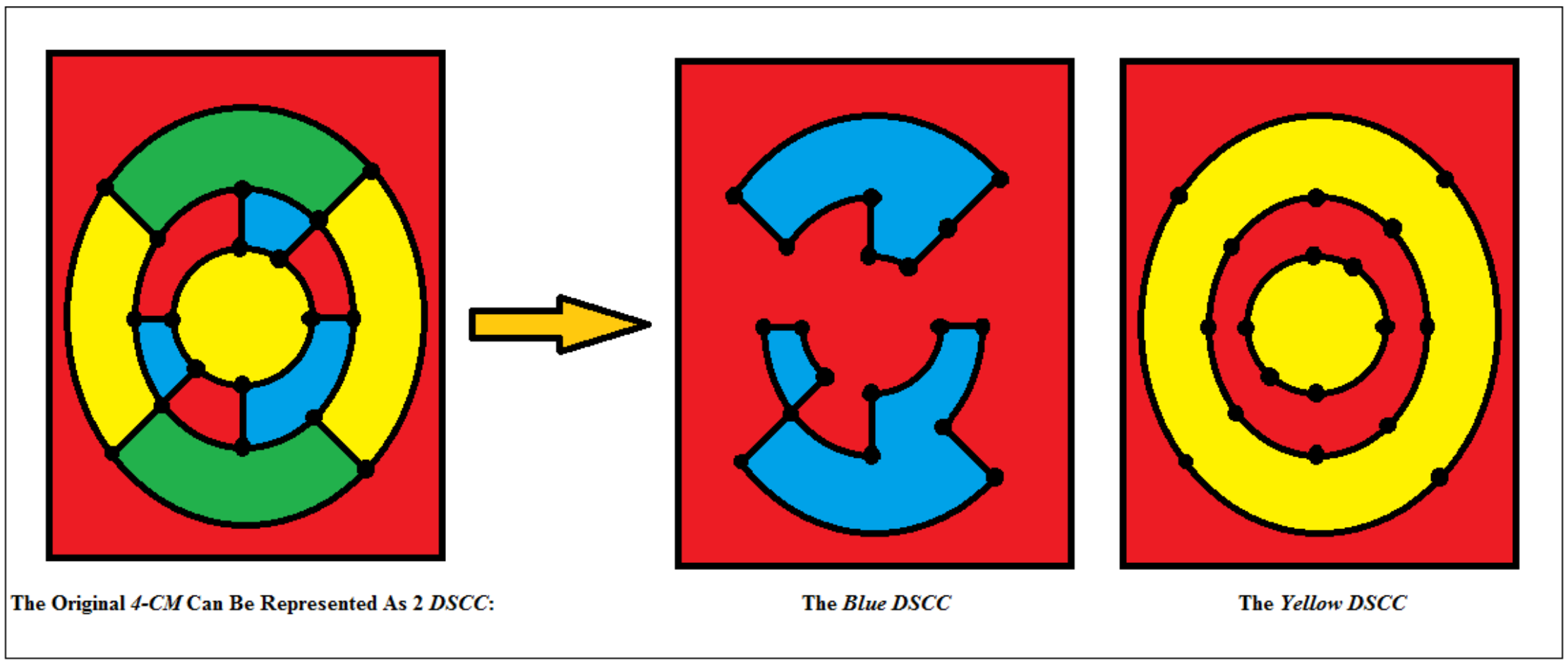

**Figure 2.3** Example 1 of How to Construct a *Blue and* a *Yellow DSCC* from a 4-CM

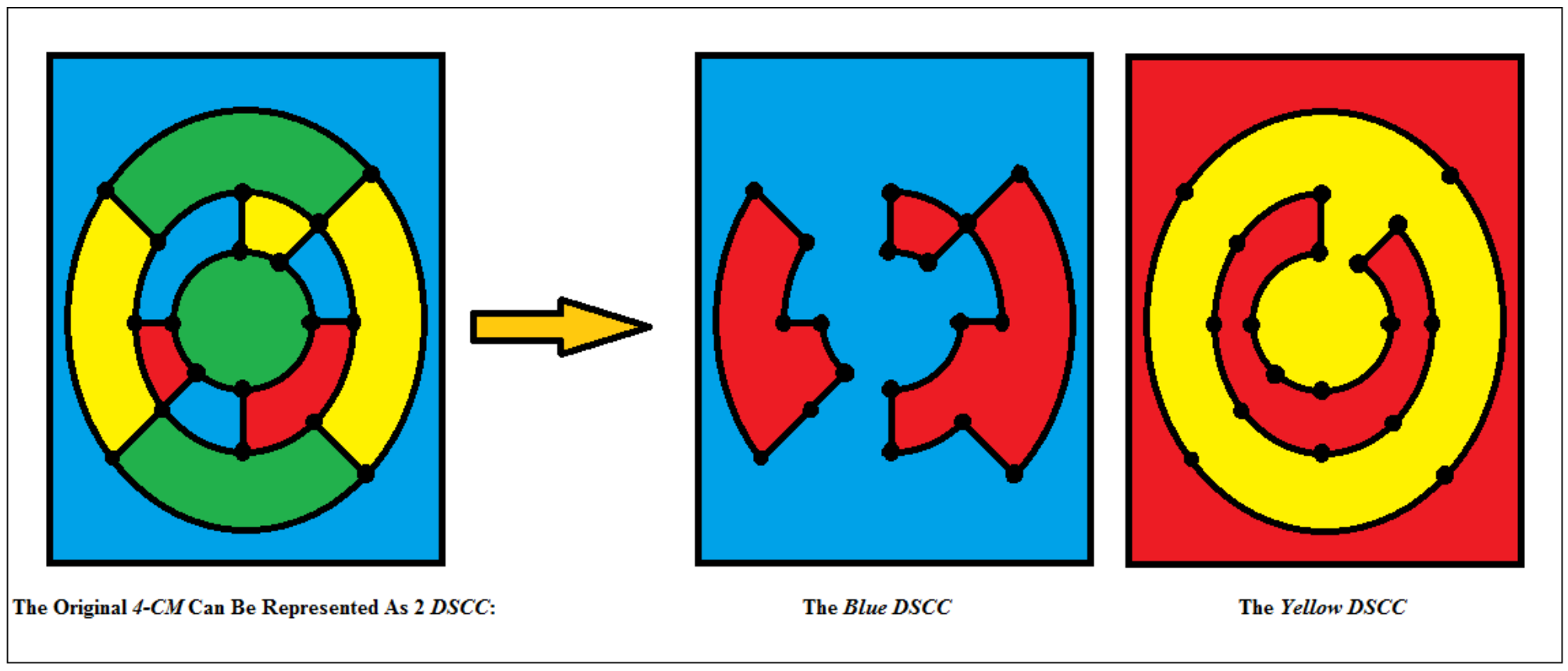

**Figure 2.4** Example 2 of How to Construct a *Blue and* a *Yellow DSCC* from a 4-CM

Hence, by the construction above, every *4-CM* is also a *2-DSCC_M*. ☐

So, by Lemmas 2.4 and 2.5, we can conclude that *2-DSCC_M* and *4-CM* are equivalent definitions, representing the same class of maps, with the same coloring properties.

**Lemma 2.6.** The quantity of blue (respect., yellow) plus green edges incident to every vertex from a *2-DSCC_M* must be even. A *2-DSCC_M* colored in this way is called *properly edge-colored*.

*Proof.* As all the curves from an arbitrary *DSCC* are closed curves, by Def. 2.1, and all green edges represent overlap of two edges (blue and yellow ones), the quantity of incident blue (respect., yellow) plus green edges on every vertex of an arbitrary 4-CM is (must be) even, for the edges in each *DSCC* always occur in pairs at the vertices, one incoming and other outgoing w.r.t. each vertex of that *2-DSCC_M*. ☐

**Corollary 2.1.** Every 3-degree vertex from a *2-DSCC_M* properly edge-colored can only have their three incident edges colored with blue, yellow and green (where the order does not matter), that is, all its vertices are properly 3-edge-colored ones.

**Definition 2.8. CBG.** A ***CBG*** is a connected finite simple planar cubic bridgeless graph (or a cubic polyhedral map). [6, 16, 17]

**Definition 2.9. 3-ECC.** A ***3-ECC*** is a *CBG* that admits a *Tait coloring* (a proper 3-edge coloring). [6, 16, 17]

**Definition 2.10. Blue-green, yellow-green and blue-yellow chain.** A *blue-green* (respect., *yellow-green* or *blue-yellow*) *chain* is a cycle in a *2-DSCC_M* that contains only blue (respect., yellow) and green (respect., yellow) edges. Note that it is allowed that there are adjacent edges with the same color in that cycle (when the map is <u>not</u> cubic).

**Definition 2.11. Local inversion of colors.** A *local inversion of colors* is swapping a color for another in all the edges in a [blue-green, yellow-green, or blue-yellow] chain.

**Lemma 2.7.** If two edges are in a blue (respect., yellow) simple cycle (closed walk without repetitions of vertices) [9] of a curve of a *DSCC* from a *2-DSCC_M* and are incident to the 5-edged vertex ***A***, where all the other vertices of this *2-DSCC_M* are 3-edged, then there is a blue-green (respect., yellow-green) chain containing these two edges in that *2-DSCC_M*.

*Proof.* As all the curves from an arbitrary *DSCC* are closed ones, and all the green edges represent overlapping of a blue and a yellow edges, then when we walk at that blue (respect., yellow) simple cycle, all the edges that we can pass by (only once, for all the others vertices are 3-edged, but ***A***) are either blue (respect., yellow) or green edge (when that blue (respect., yellow) curve intersects a yellow (respect., blue) one at infinitely many points), that is, we are walking only upon blue (respect., yellow) and green edges, where this closed walking forms a blue-green (respect., yellow-green) chain, starting in ***A*** and returning to it. ☐

Notice that in general the blue-green (respect., yellow-green) chains herein can have adjacent edges with the same color, by Def. 2.4, but in a 3-edge properly colored *3-ECC* this is <u>not</u> possible, where every adjacent edges must have alternating color, since otherwise would be two edges with the same color adjacent to some properly 3-edged-colored vertex, which is not possible, by Corollary 2.1.

**Lemma 2.8.** Every ***3-ECC*** is also a ***4-CM*****.**

*Proof.* By *CBG*'s Tait coloring, every *3-ECC* is also a *4-CM* [10, 16, 17]. ☐

See that we can get a *2-DSCC_M* from a *3-ECC* simply considering all its blue-green (respect., yellow-green) chains as blue (respect., yellow) curves, so obtaining a *2-DSCC_M*.

**Corollary 2.2.** Every ***3-ECC*** is also a ***2-DSCC_M*** (***3-ECC*** ⇒ ***4-CM*** ⇔ ***2-DSCC_M***, hence ***3-ECC*** ⇒ ***2-DSCC_M***), by Lemmas 2.4, 2.5 and 2.8.

**Theorem 2.1. Four-Color Theorem (4CT).** Every connected finite simple planar graph is a ***4-CM***. [1, 16, 17]

*Proof.* As very well known, 4CT is equivalent to the Theorem 2.2 below [7, 11, 16, 17]:

**Theorem 2.2. Three Edge-Coloring Theorem (3-ECT).** Every *CBG* is a *3-ECC*.

Therefore, we shall prove the 3-ECT, hence the 4CT. So, as it is well known too, since every finite map that has no region completely surrounded by another region can be converted into a cubic map, an arbitrary hypothetical *minimal CBG* that is a counter-example to the 4CT (and to the 3-ECT too, naturally) would be a *CBG* ***N*** that has a pentagon inside it. [2, 17]

Thus, the resultant map ***C*** when that pentagon is removed from ***N*** is a *4-CM*, since ***N*** is hypothetically *minimal*, as shown in the sequence of Figs. 2.5, 2.6 and 2.7 below (for if ***C'*** in the Figure 2.5 below [***C'*** = ***N*** without an arbitrary edge from that pentagon] was not a four-colorable map, then it would be a *CBG* ***C'*** that is a counter-example smaller than ***N***, which is impossible, as ***N*** is a <u>*minimal*</u> one, by hypothesis); so, ***C*** is also a *2-DSCC_M*, by Lemma 2.5:

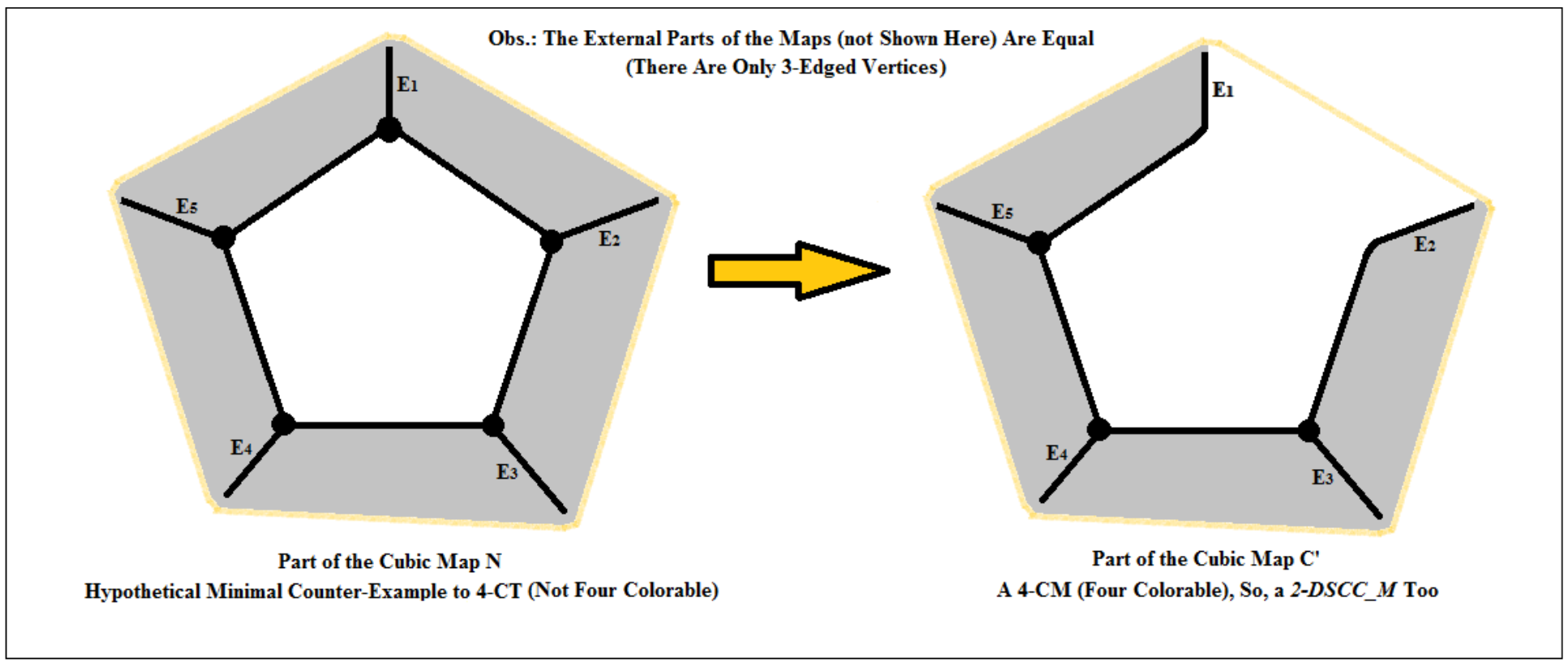

**Figure 2.5** A hypothetical minimal *CBG* that is a counter-example to the 4CT can originate a smaller *3-ECC*

Verify above that, as ***N*** is a cubic map, then ***C'*** is a cubic map too:

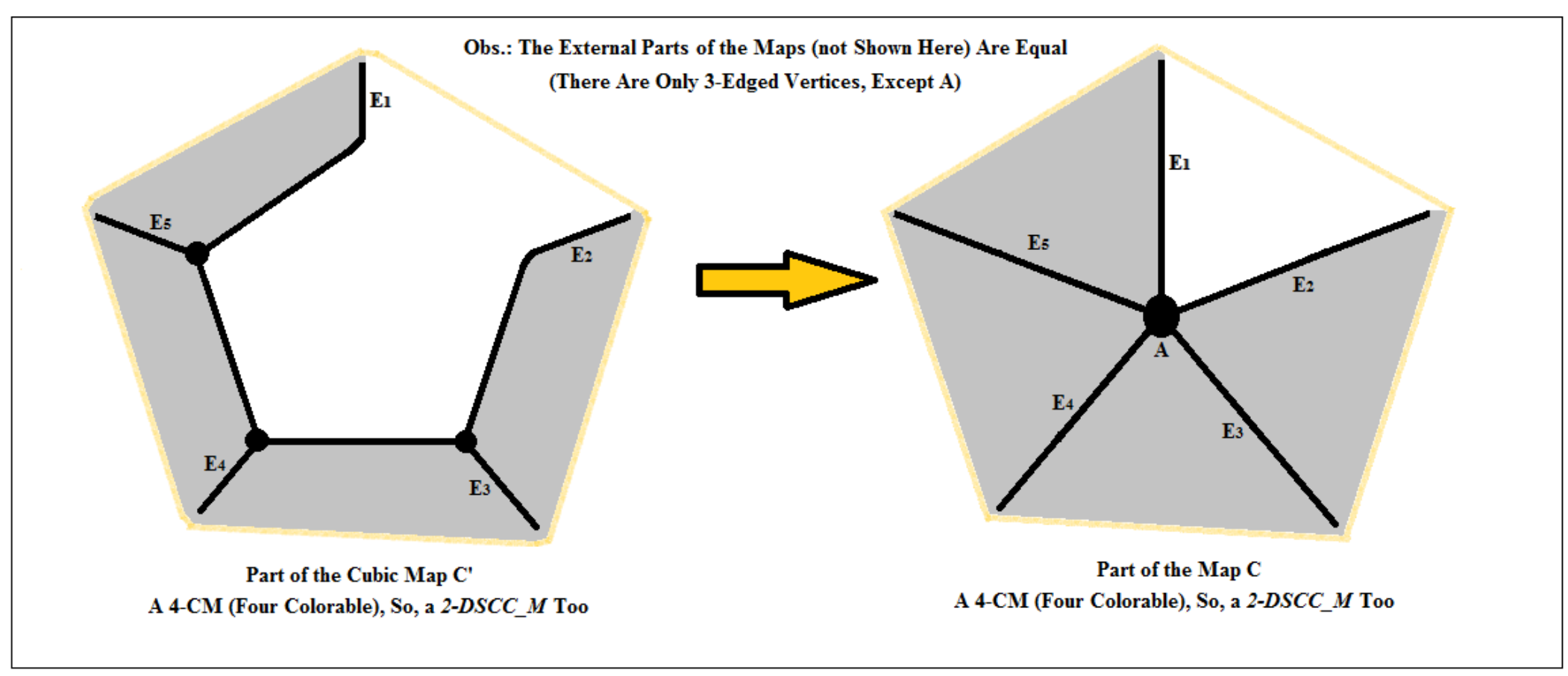

**Figure 2.6** Demonstration that from the *3-ECC **C'*** we can create the *2-DSCC_M **C***

Then, joining the two previous figures, we reach the result seen in the Fig. 2.7 below, where from that map ***N*** we construct that *2-DSCC_M **C*** (that is not a cubic map, unlike of that *3-ECC **C'***), with that pentagonal face contracted:

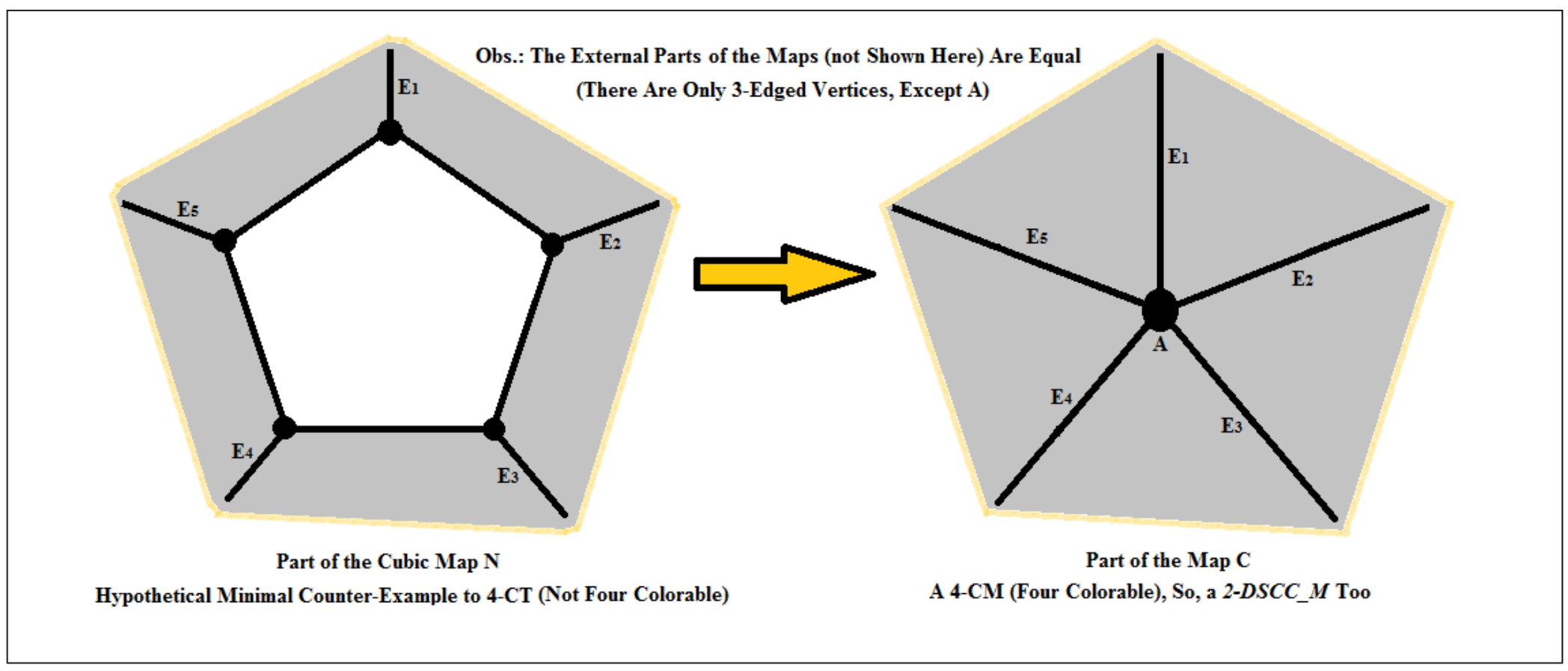

**Figure 2.7** Demonstration that from a hypothetical minimal counter-example we can create a *2-DSCC_M*

In the resultant *2-DSCC_M **C*** at the right in the Fig. 2.7 (note that a *2-DSCC_M* does not need to be cubic, by Def. 2.5), the quantity of blue (respect., yellow) plus green edges incident to the vertex **A** must be even (by the Lemma 2.6), where the only quantities allowed are 3-1-1 (one color appears three times and the others only once), as demonstrated by exhaustion in the table below, where **σ** represents an arbitrary permutation of those five edges:

| $E_{\sigma(1)}$ | $E_{\sigma(2)}$ | $E_{\sigma(3)}$ | $E_{\sigma(4)}$ | $E_{\sigma(5)}$ | Qty. Blue | Qty. Yellow | Allowed? |
|---|---|---|---|---|---|---|---|
| Blue | Blue | Blue | Blue | Blue | Odd (5) | Even (0) | No |
| Blue | Blue | Blue | Blue | Yellow | Even (4) | Odd (1) | No |
| … | … | … | … | … | … | … | … No … |
| … | … | … | … | … | … | … | … No … |
| Blue | Blue | Blue | Yellow | Green | Even (4) | Even (2) | Yes |
| Blue | Yellow | Yellow | Yellow | Green | Even (2) | Even (4) | Yes |
| Blue | Yellow | Green | Green | Green | Even (4) | Even (4) | Yes |

**Table 2.4** Quantities Allowed of Colored-Edges Incidents to the Vertex ***A*** of the Map ***C*** in the Fig. 2.8

We shall analyze only the case with three blue edges, one green and one yellow, since the other cases are only permutations of those colors, leading to the same results, by symmetry (rotation or reflection of the map, or global inversion of some pair of colors).

**Definition 2.12. TBCI Map.** A *TBCI map* is as that ***C*** shown in the Fig. 2.8, where those three blue edges are contiguously incident to the vertex ***A*** (that is, we can draw a continuous line crossing all these three edges without crossing any other edge from that map).

This definition is general, embracing all the possible cases where three edges with the same color are contiguously incident to that vertex ***A***, up to the colors and their positioning. Then, in a *TBCI map* we can create a new region expanding that vertex ***A***, returning to the original map ***N***, but now proving that it is really a *3-ECC*, as shown in the Fig. 2.8, so that map ***N*** is really a *4-CM* too, by Lemma 2.8, hence in this case it cannot be actually a minimal counter-example to the 4CT. We shall see more details below:

Well, as all the vertices from that map ***C***, except the vertex ***A***, are properly 3-edge-colored, by Corollary 2.1, then when we include that new region expanding the vertex ***A***, we demonstrate that that map ***N*** is also a *3-ECC*, as the five new vertices replacing the vertex ***A*** are also all properly 3-edge-colored (see it at Fig. 2.8), and all the remaining vertices of ***N*** are so too, because the external part of ***N*** is equal to the external part of ***C*** (even though they are not explicitly shown in the figure below):

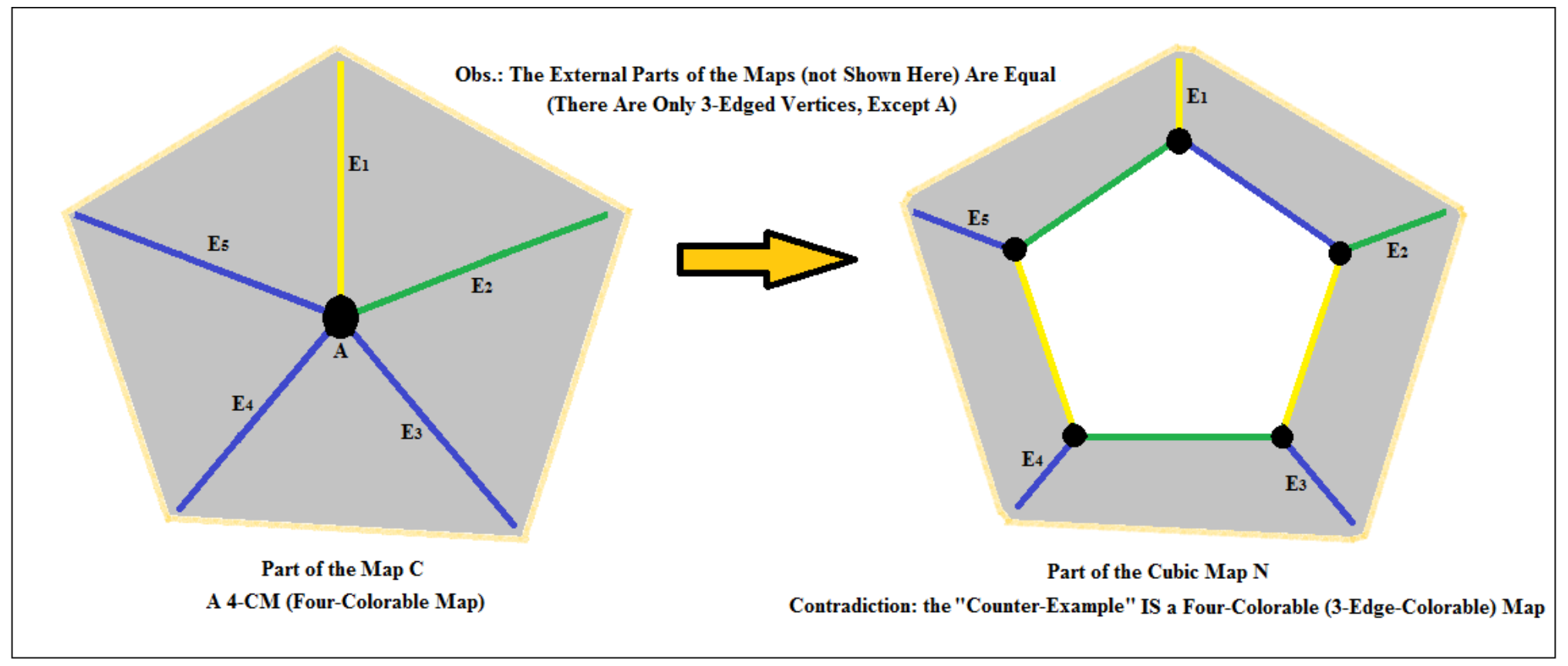

**Figure 2.8** Demonstration that a supposed minimal counter-example (***N***) is NOT really a counter-example

So, since we can delete a region from the map ***N***, then properly color that resultant map ***C*** with only four colors, and then return that region to ***N***, generating a *3-ECC*, so also a *4-CM*, by Corollary 2.2, this process proves that the original map ***N*** cannot be a true minimal counter-example to the 4CT, which yields a contradiction, for our hypothesis is that ***N*** is so. Consequently, the initial assumption that ***C*** is a *TBCI map* is (must be) wrong, in order to maintain our hypothesis (even though this is temporary, as we'll see it below). Therefore, that *4-CM* map ***C*** is <u>not</u> a *TBCI map*; hence it must be like ones shown in the Fig. 2.9:

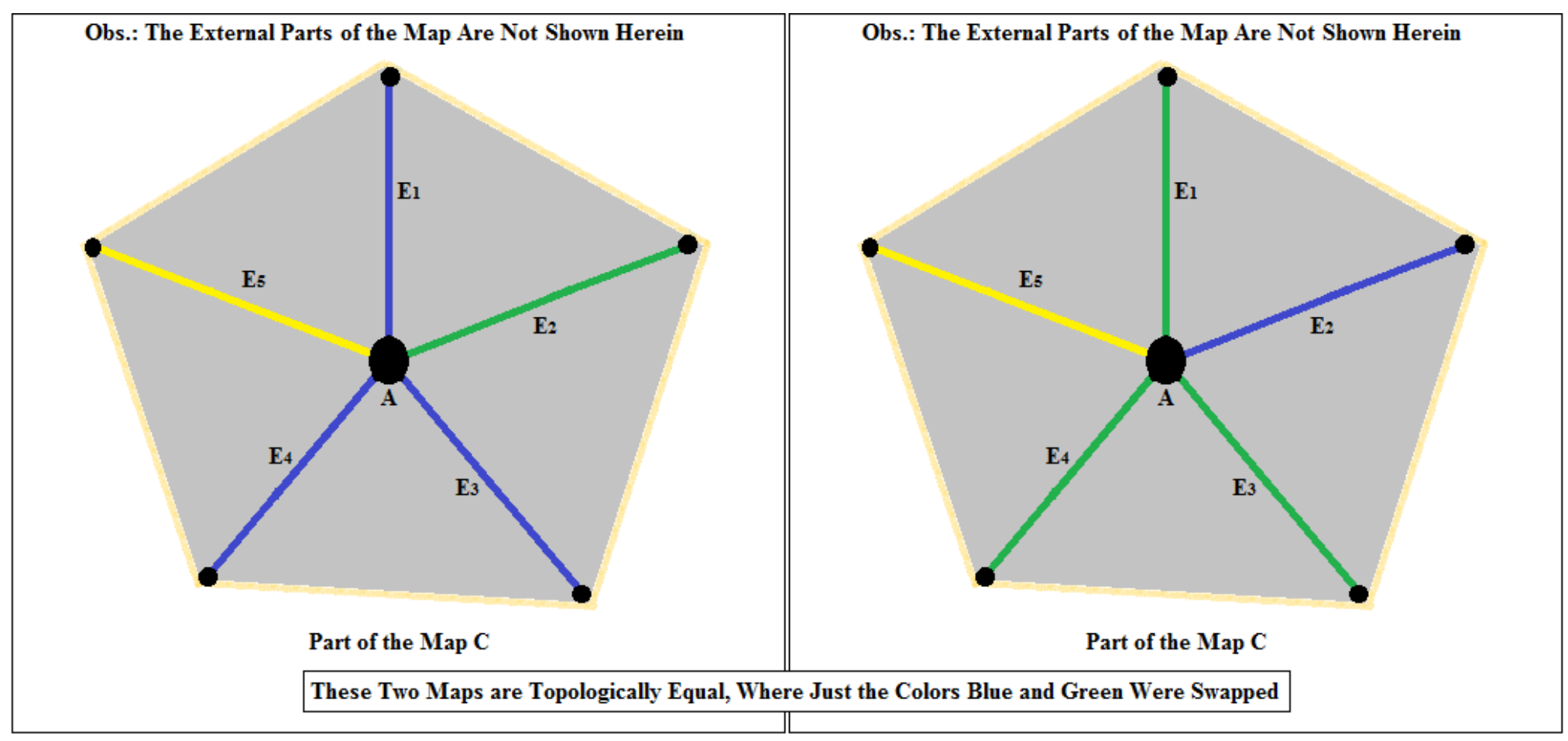

**Figure 2.9** In order to ***N*** can be a minimal counter-example to the 4CT, ***C*** must have this kind of coloring

So, two possible topologies (structures or conformations) of the blue curve from that map ***C*** (where the edges $E_1$, $E_3$ and $E_4$ are blue, $E_2$ is green, and $E_5$ is yellow [alternatively, if $E_1$, $E_3$ and $E_4$ are green (respect., yellow), $E_2$ is blue (respect., green), and $E_5$ is yellow (respect., blue), we swap the colors blue and green (respect., yellow), generating a map whose topological analyses would be identical to that one we shall do w.r.t. that map ***C***]) are represented in the Fig. 2.10, since those blue edges must belong to the blue curve(s), and that (those) blue curve(s) must pass by the vertex ***A*** and by all the other five vertices surrounding it, and by those four edges: three blue ($E_1$, $E_3$ and $E_4$) and one green ($E_2$) (note that a green edge is the result from overlapping blue and yellow ones):

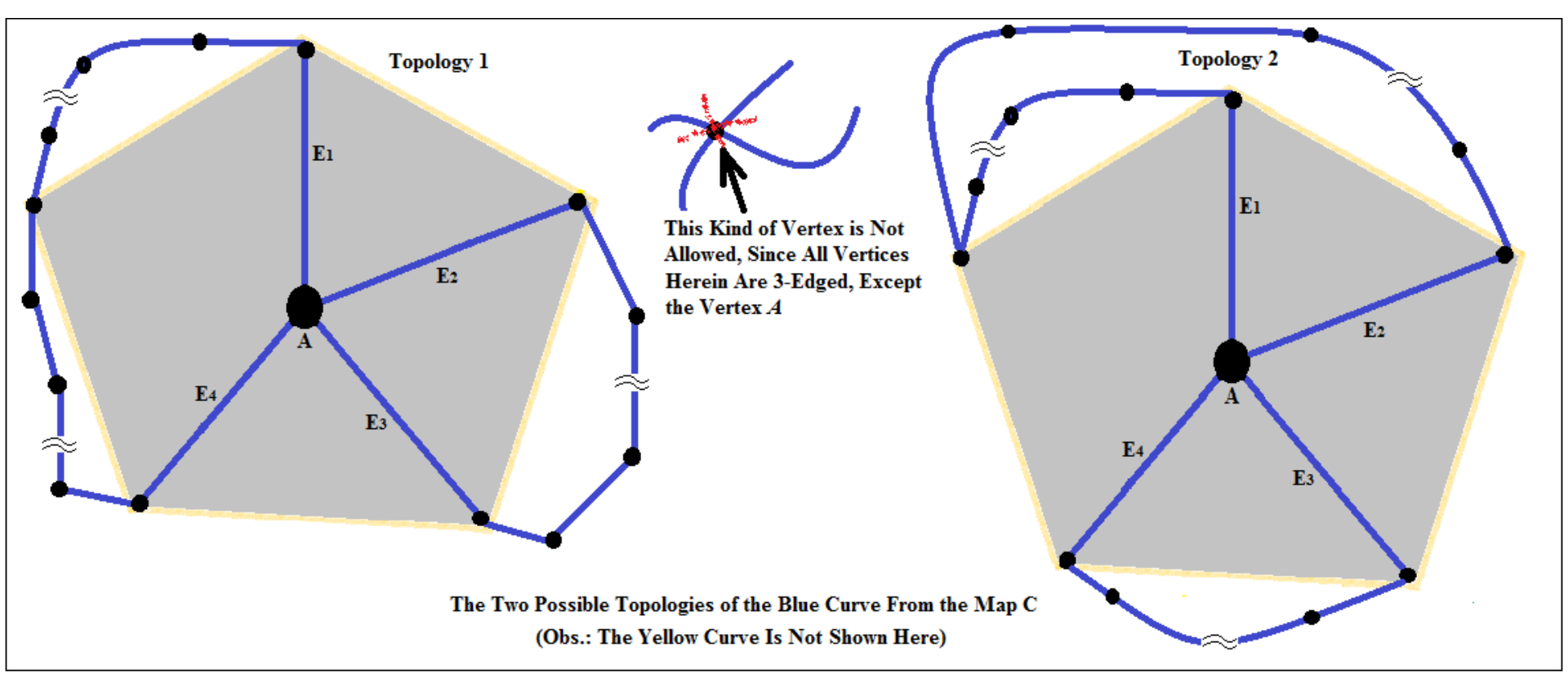

**Figure 2.10** Two possible fit topologies of the blue curve that pass by the vertex ***A*** from the map ***C***

Note that there do exist more two other possible topologies, 1' and 2', but they are essentially the same as those 1 and 2 represented in the Fig. 2.10, respectively, since all following conclusions are completely applicable to them too, when we replace the topology 1 by the 1', or the topology 2 by the 2' (obs.: verify by exhaustion that there is no other ones):

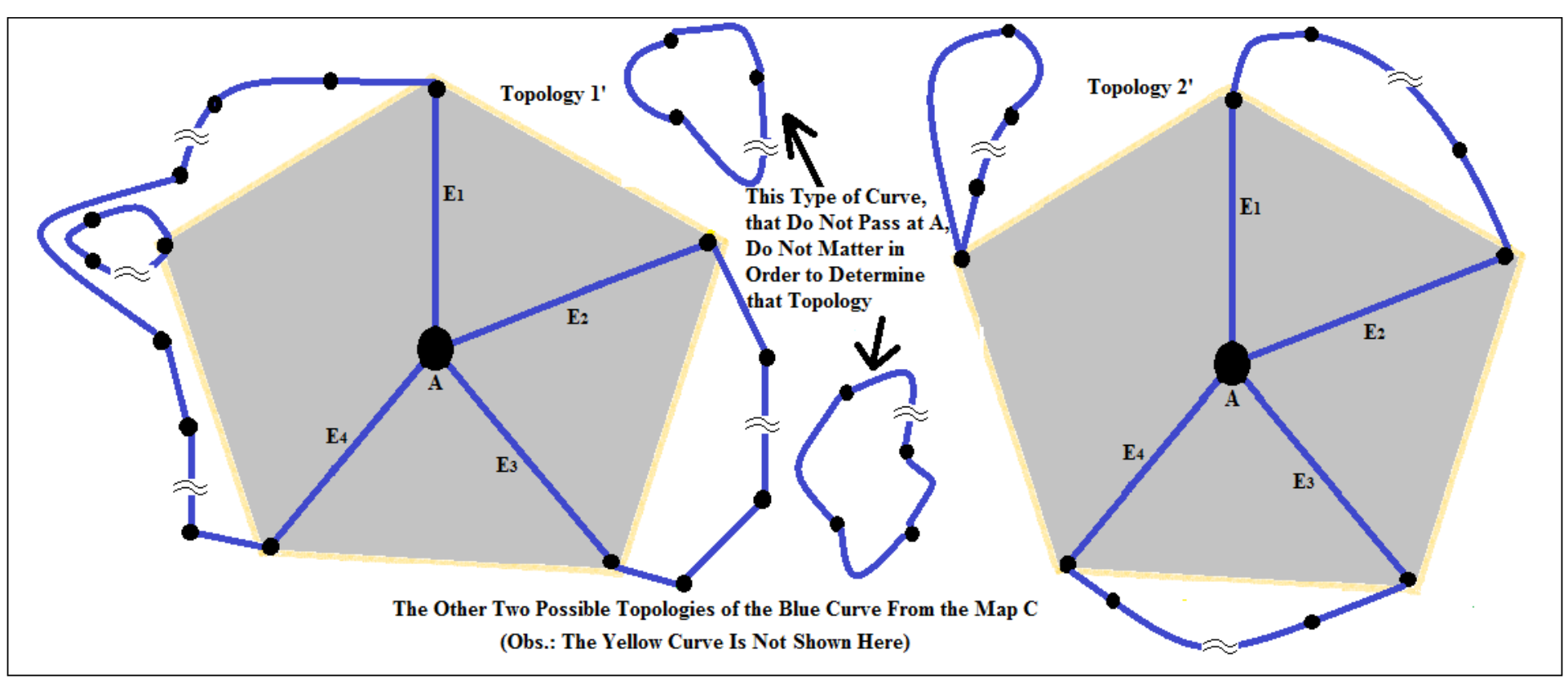

**Figure 2.11** Other two possible topologies of the blue curves that pass by the vertex ***A*** of the map ***C***

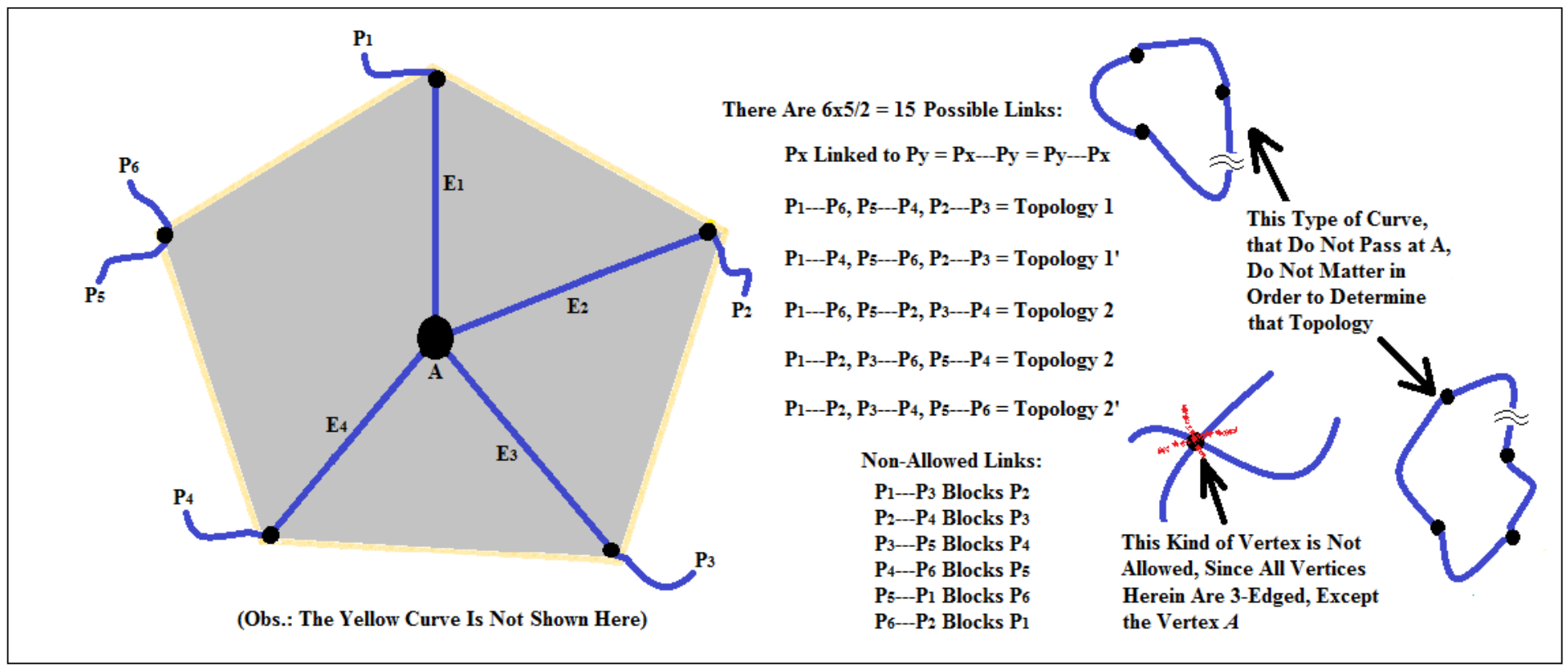

**Figure 2.12** Demonstration that there are herein only these four possible topologies: 1, 1', 2 and 2'

Note: Furthermore, other blue curves that do not pass by any of these six vertices can exist, but these curves, even though they really there do exist, do not matter at all w.r.t. the topologies above and the proof in this paper, since the arguments herein utilized shall be demonstrated true ones independently of the existence or nonexistence of such curves.

Verify yet that there is no other possible fit topologies for any blue curve (or curves) that pass by all those six vertices and four edges: it is enough to think of a torn blue curve with loose ends at those six vertices and four edges and exhaustively try to link those loose ends ($P_1$, $P_2$, $P_3$, $P_4$, $P_5$ and $P_6$ in the Fig. 2.12) of the curve in order to fix the entire curve into a repaired closed one, with neither intersection nor crossing of it with itself (nor with another curve of the same color), except at vertex ***A***, as shown in the Fig. 2.12.

Notice that there do exist two possible topologies to the map ***C***, but if it had topology 2 (or 2'), however, then would exist a blue-green chain including $E_1$ and $E_2$ (by Lemma 2.7, since these edges are contained in a simple cycle, as shown in the Fig. 2.10 – and in the Fig. 2.11, w.r.t. topology 2'), which would permit a local inversion of these colors in this chain (as in Def. 2.11), generating a 3-edge-colored map with three blue edges contiguously incident to the vertex ***A***, allowing for the creation of more one region into ***C*** and it continuing being properly 3-edge-colored (as in the Figure 2.8), which also would imply that that map ***N*** could not be a true minimal counter-example to the 4CT, as shown in the figure below:

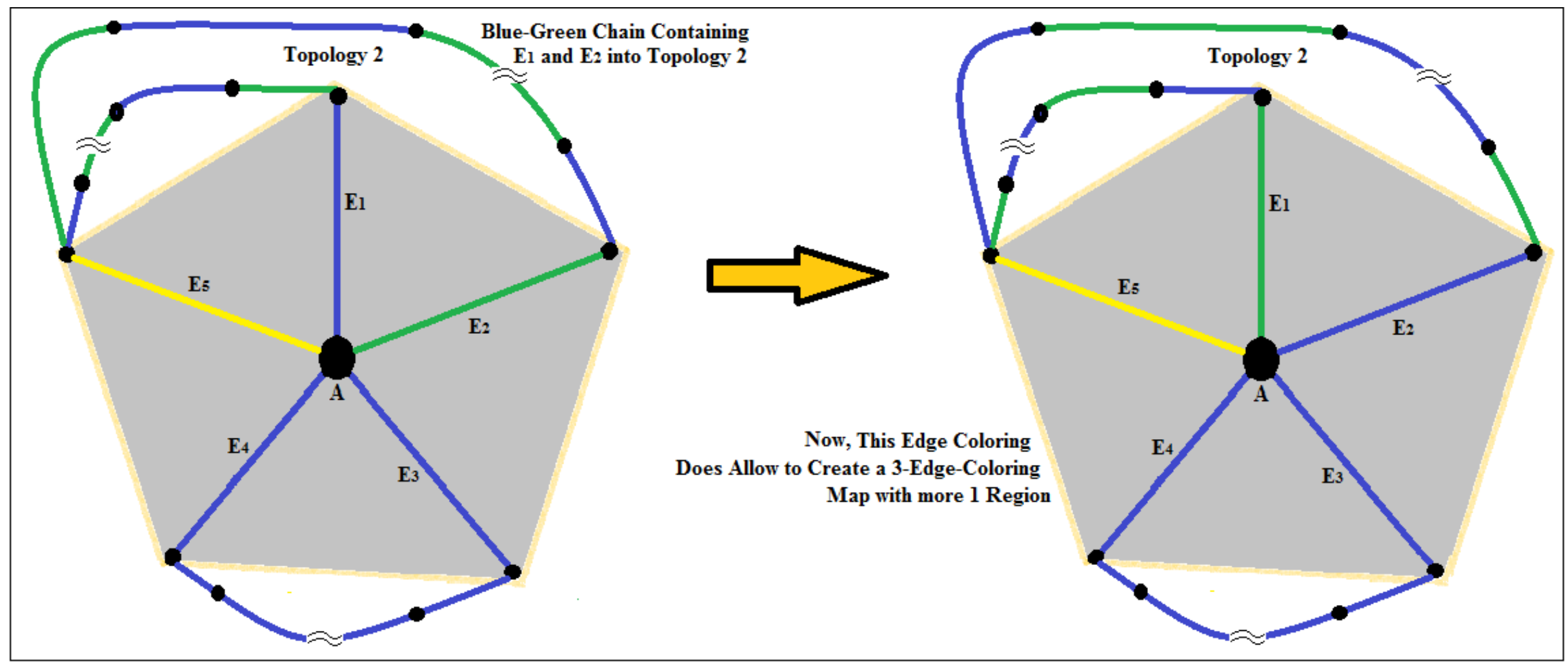

**Figure 2.13** Locally inverting colors in a blue-green chain, then resulting a *TBCI map*

Hence, as by hypothesis that map ***N*** should be a minimal counter-example to the 4CT, that topology is (must be) the 1 (or the 1').

So, as the three blue edges ($E_1$, $E_3$ and $E_4$) are not contiguously incident to the vertex ***A***, and they are positioned according to topology 1 (or 1'), we can locally invert the colors blue and yellow in a blue-yellow chain containing $E_1$ following exactly one from the three ways below:

1. The blue-yellow chain contains the edges $E_1$ and $E_5$, as shown in the Fig. 2.14, where that local inversion of colors generates a *TBCI map*, which implies that same conclusion (that map ***N*** cannot be a real minimal counter-example to the 4CT), and then producing that same contradiction with our hypothesis (that that map ***N*** is a true minimal counter-example to the 4CT), as shown in the figure below:

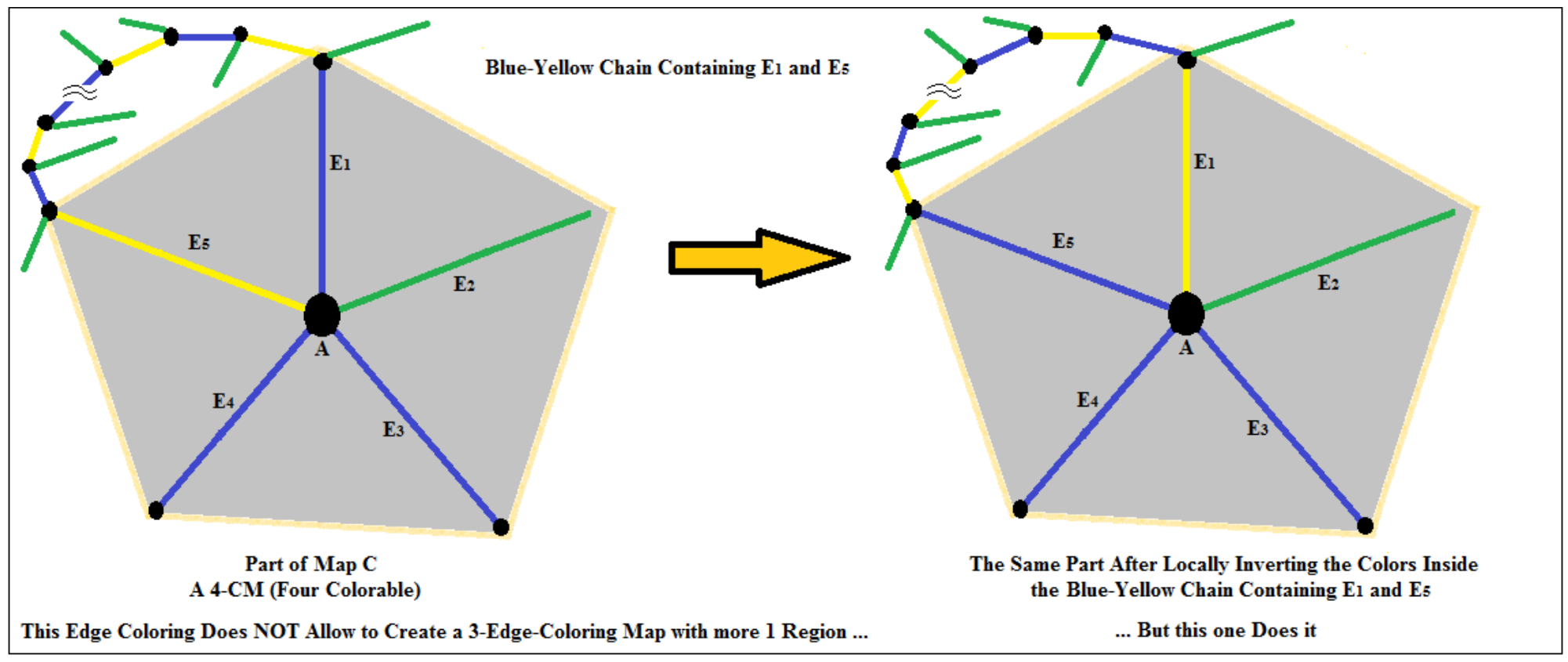

**Figure 2.14** Locally inverting colors in a blue-green chain, then resulting a *TBCI map*

2. The blue-yellow chain cannot contain the edges $E_1$ and $E_4$, for if so, then it would block the blue-yellow chain containing the edges $E_3$ and $E_5$, by planarity, as shown in the Fig. 2.15, where that local inversion of colors in this chain which also would generate a *TBCI map* cannot really occur:

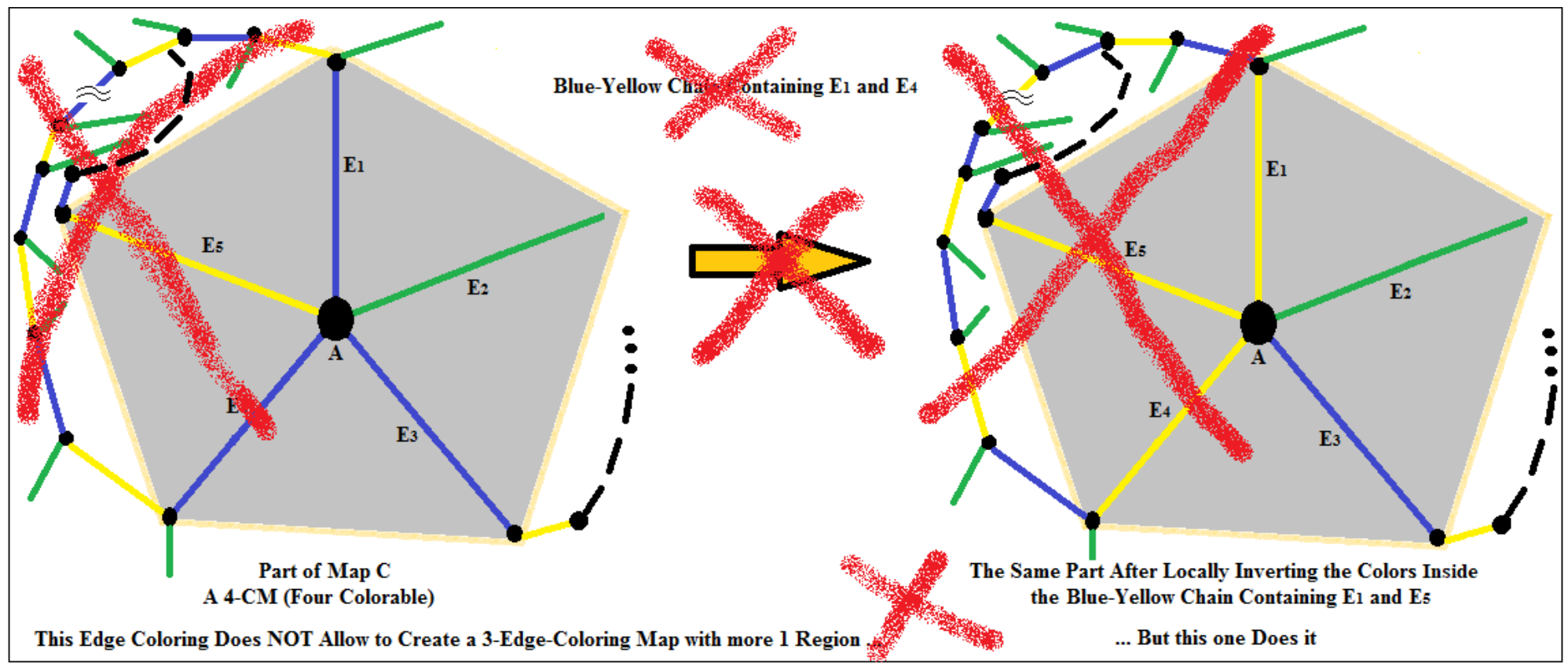

**Figure 2.15** These blue-yellow chain containing the edges $E_1$ and $E_4$ and local inversion cannot occur

3. The blue-yellow chain contains the edges $E_1$ and $E_3$, as shown in the Fig. 2.16. Now, however, that local inversion of colors generates a map that is not a *TBCI map*, whereby that conclusion (our hypothesis is false) is no more valid here (that map ***N*** now seems right, for it can still be a minimal counter-example to the 4CT).

**Definition 2.13. Local inversion *L₁*.** That local inversion of colors that does not generate a map ***C*** locally colored at vertex ***A*** in a way like the one in the Fig. 2.8 (a *TBCI map*), and that passes by two noncontiguous edges incident to ***A***, is defined as *Local Inversion* $L_1$, as shown in the figure below:

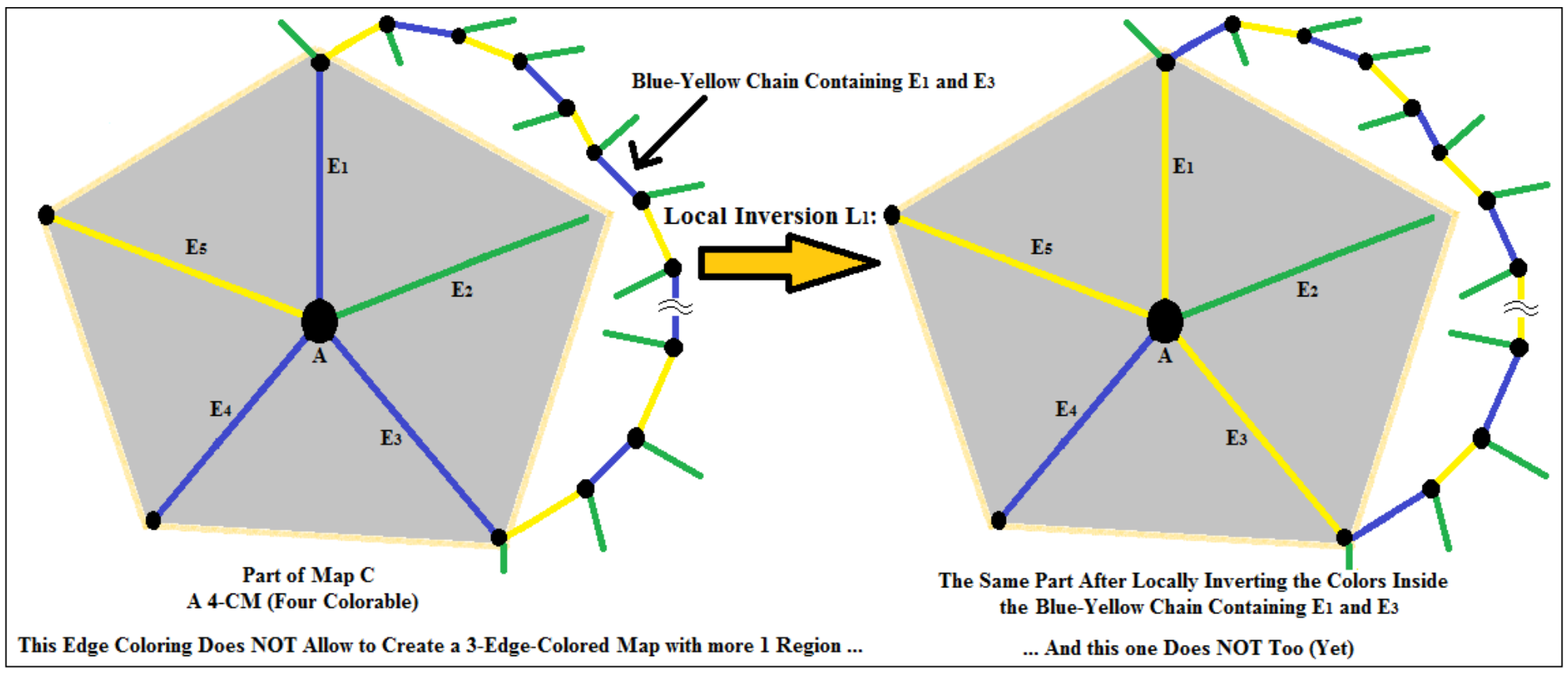

**Figure 2.16** Locally inverting colors in a yellow-green chain resulting a map that is not a *TBCI map* too

**Proposition 2.2.** A *Local Inversion* $\boldsymbol{L_1}$ (any arbitrary local inversion, really) leaves the property of being *2-DSCC_M* intact about the map ***C*** above, that is, after it that map ***C*** continues being a *2-DSCC_M*.

*Proof.* After that local inversion of colors $L_1$ (or any other arbitrary local inversion), all the 3-edged vertices of that map ***C*** continues properly colored, because there was only an exchanging of the colors in two incident edges in each 3-edged vertex. And the 5-edged vertex ***A*** can be expanded into three properly colored 3-edged vertices ($A_1$, $A_2$, and $A_3$), as shown in the Fig. 2.17, generating after all a *3-ECC*, as shown in the Figs. 2.18 and 2.19; thereby the resultant map is a *2-DSCC_M* too, by Corollary 2.2. Then, contracting those three vertices ($A_1$, $A_2$, and $A_3$) and returning them into the vertex ***A***, two things can occur, as shown in the Figs. 2.18 and 2.19, respectively: Either 1) the simple closed yellow curve that passes by those three vertices intersects itself at a single point (the vertex ***A***) and changes itself into a non-simple closed yellow curve; or 2) the two simple closed yellow curves that pass by those three vertices that join themselves into only one non-simple closed yellow curve. Where the simple closed blue curve that passes by those three vertices continues unaltered passing by the vertex ***A***, thereby leaving the map ***C*** continuing to be a *2-DSCC_M*, as shown in the Figs. 2.17, 2.18 and 2.19: ☐

Obs.: The left part of map below is the same as one at right in the Fig. 2.16, just drawn 144° (0.8 π rad) counter-clockwise rotated:

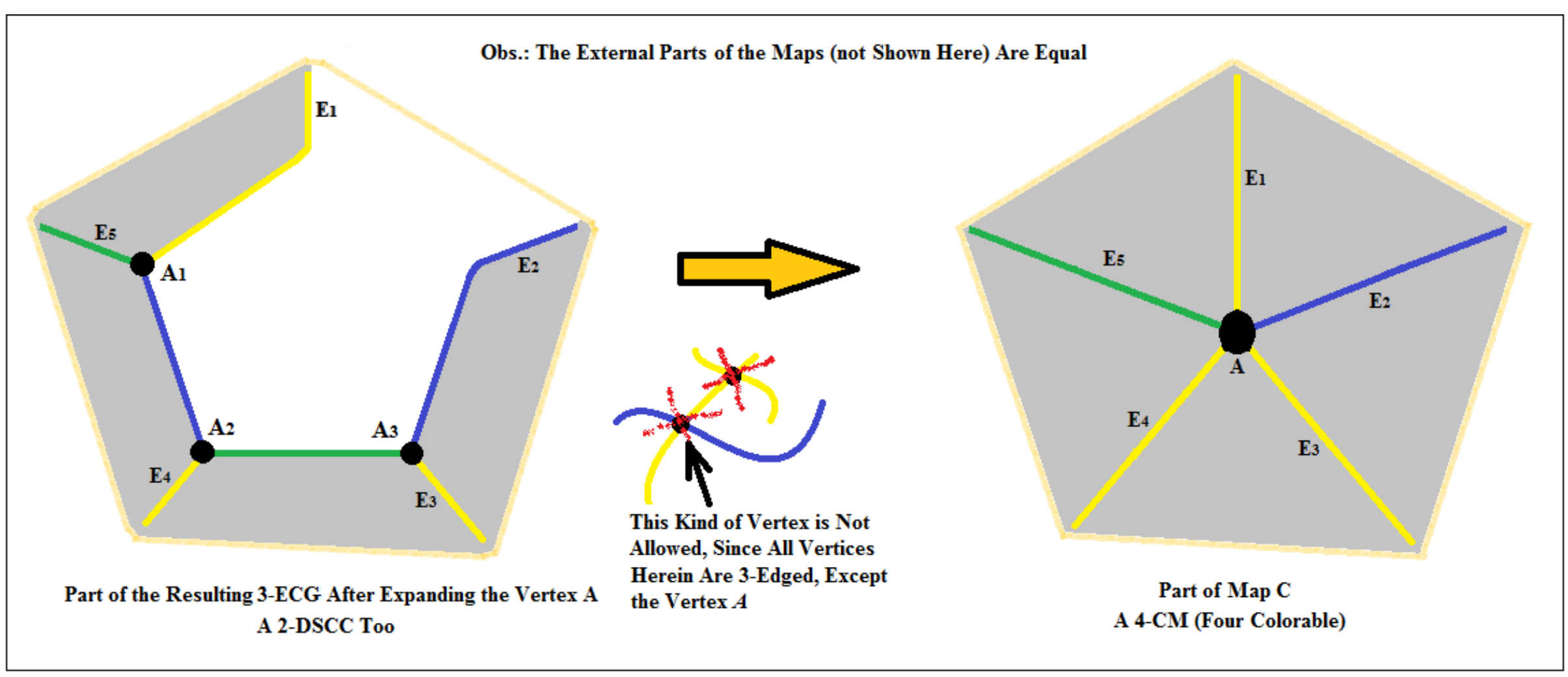

**Figure 2.17** Expanding the vertex ***A*** and generating a *2-DSCC*

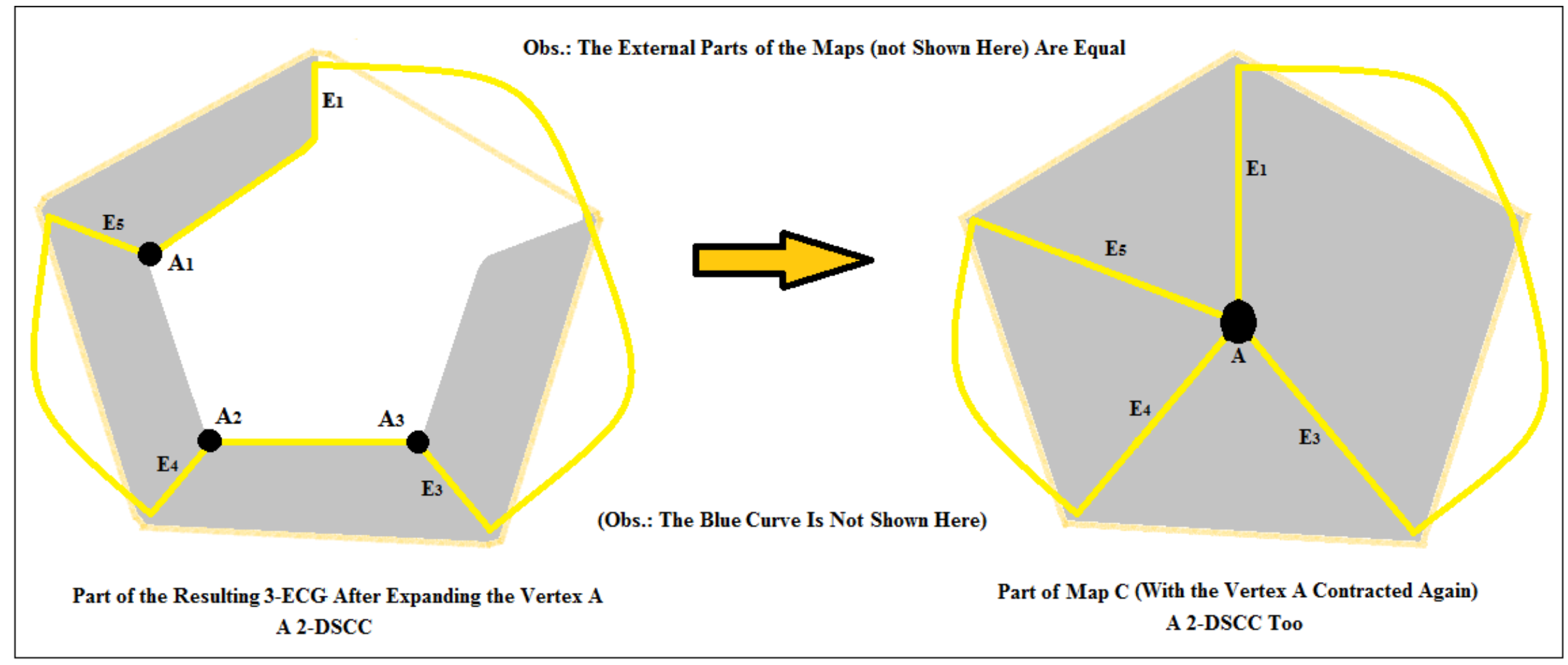

**Figure 2.18** Contracting the vertex ***A*** again and showing that the map continues being a *2-DSCC*

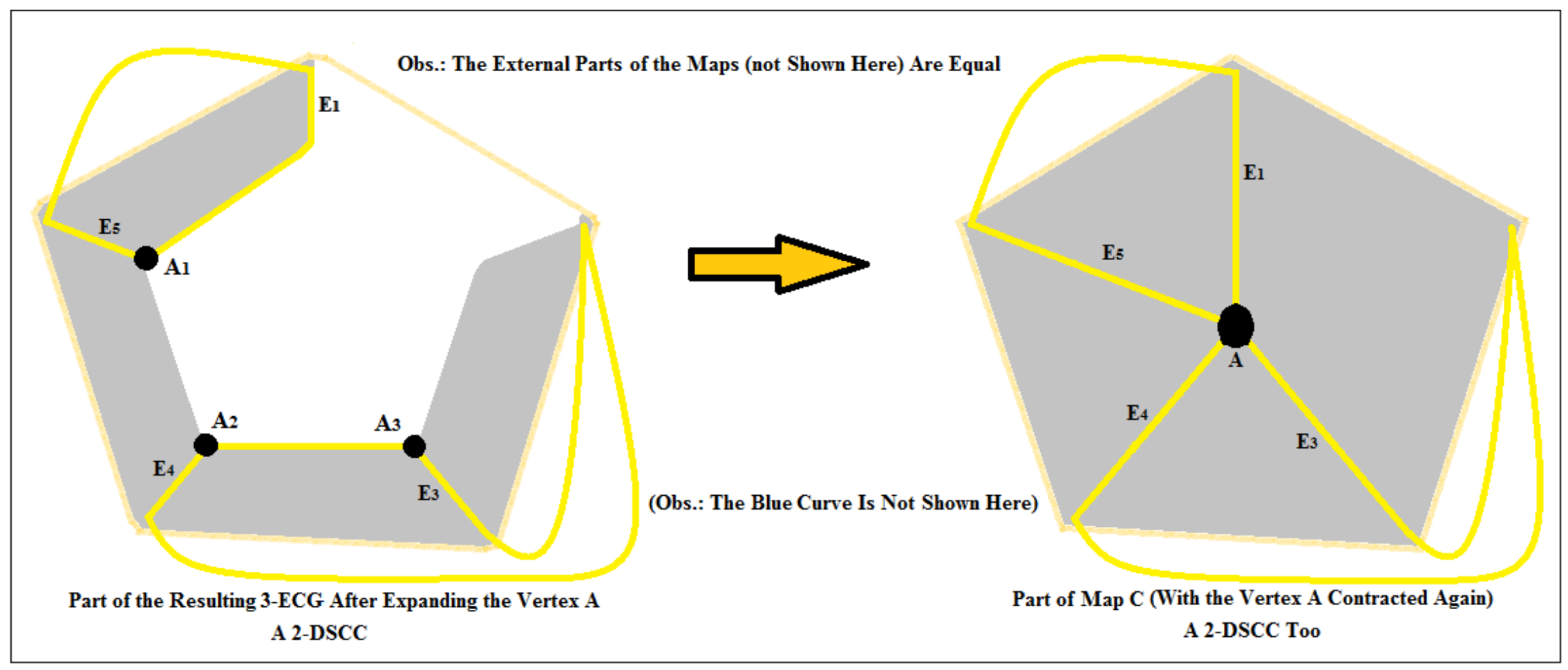

**Figure 2.19** Contracting the vertex ***A*** again and showing that the map continues being a *2-DSCC*

Consequently, the two only possible fit topologies of the yellow curve from the resultant map ***C*** above, the only acceptable ones (where the edges $E_1$, $E_3$ and $E_5$ are yellow, and $E_2$ is green), are (must be) those ones shown in the figure below:

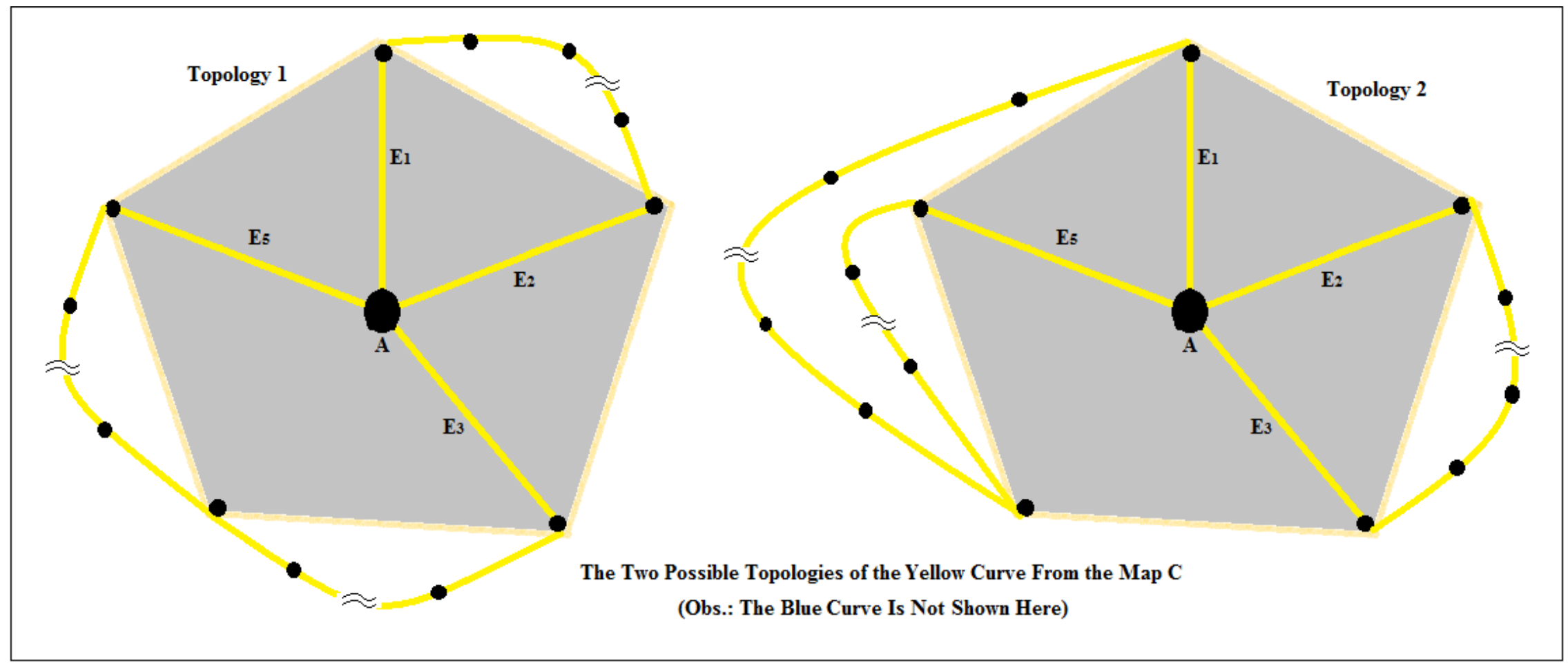

**Figure 2.20** Two possible fit topologies of the yellow curve of the map ***C*** that pass by vertex ***A***, after $\boldsymbol{L_1}$

Note yet that, as before w.r.t. that blue curves in the Fig. 2.10, here also there are two other possible fit topologies, 1' and 2'; but they too are essentially the same as those ones 1 and 2, respectively, represented in the Fig. 2.20, for the same reasons explained after the Fig. 2.10, as shown in the figure below:

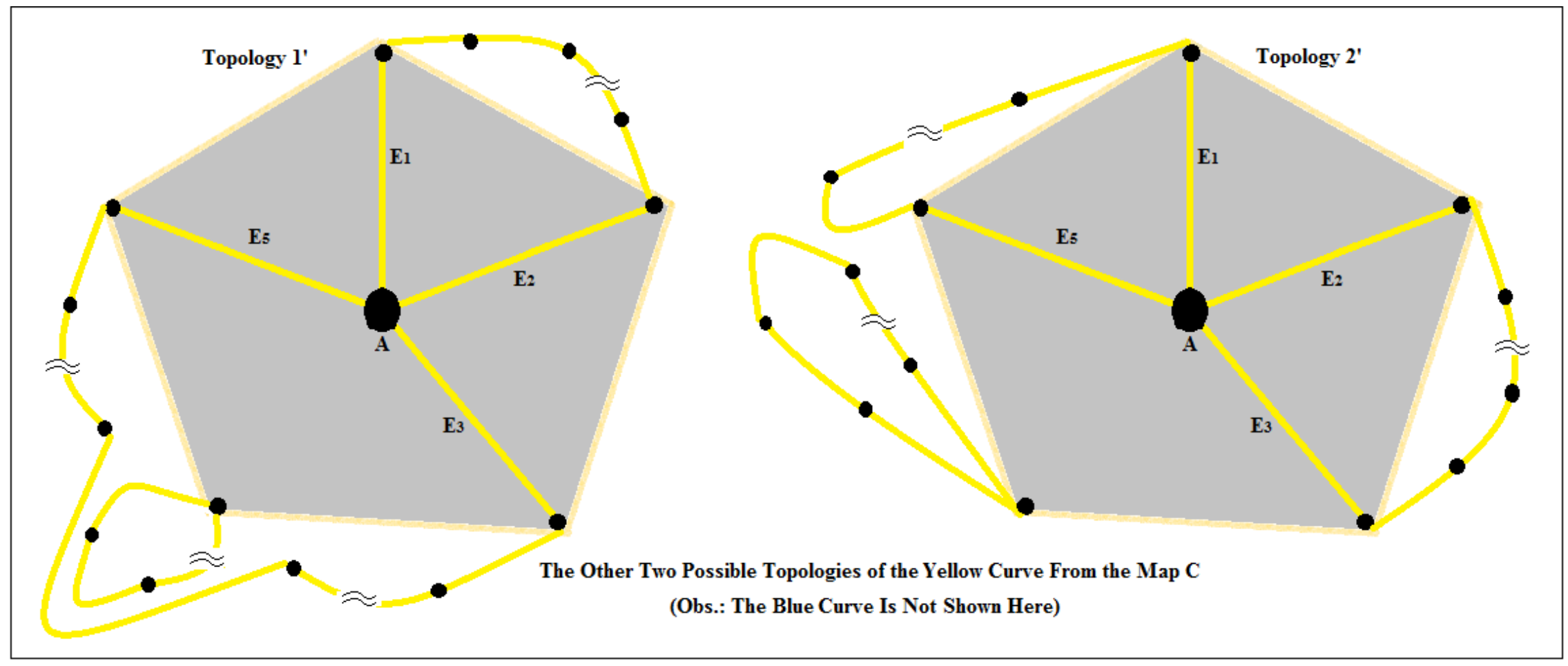

**Figure 2.21** Other two possible fit topologies of the yellow curve of the map ***C*** that pass by vertex ***A***, after $\boldsymbol{L_1}$

If it was topology 2 (or 2'), however, then, as in the Fig. 2.13, it will be a flaw too:

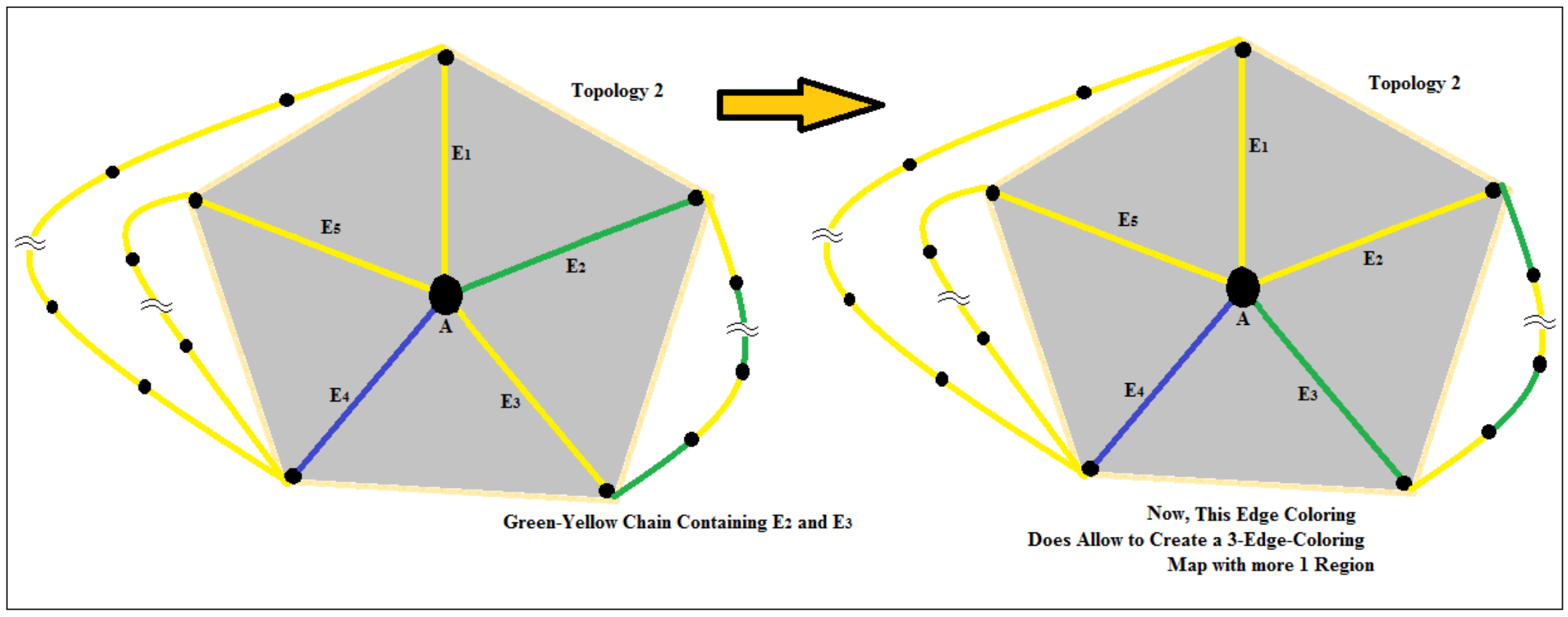

**Figure 2.22** Locally inverting colors in a yellow-green chain, then resulting a *TBCI map*

Hence, as ***N*** is a minimal counter-example to the 4CT, by hypothesis, in order to avoid that flaw above (resulting a *TBCI map* after that local inversion of colors in that green-yellow chain), that topology is (must be) also the 1 (or the 1'). Now, w.r.t. the resultant map ***C*** colored as in the way 3 above (as that map placed at right in the Fig. 2.16), we can locally invert the colors from the blue-yellow chain containing $E_4$ and $E_5$.

**Definition 2.14. Local inversion *L*₂.** The local inversion above, that generates a map ***C*** locally colored at vertex ***A*** in a way like the one in the Fig. 2.9 (which is <u>not</u> a *TBCI map*), and passes by two contiguous edges incident to ***A***, is defined as *Local Inversion* $\boldsymbol{L_2}$, as shown in the figure below:

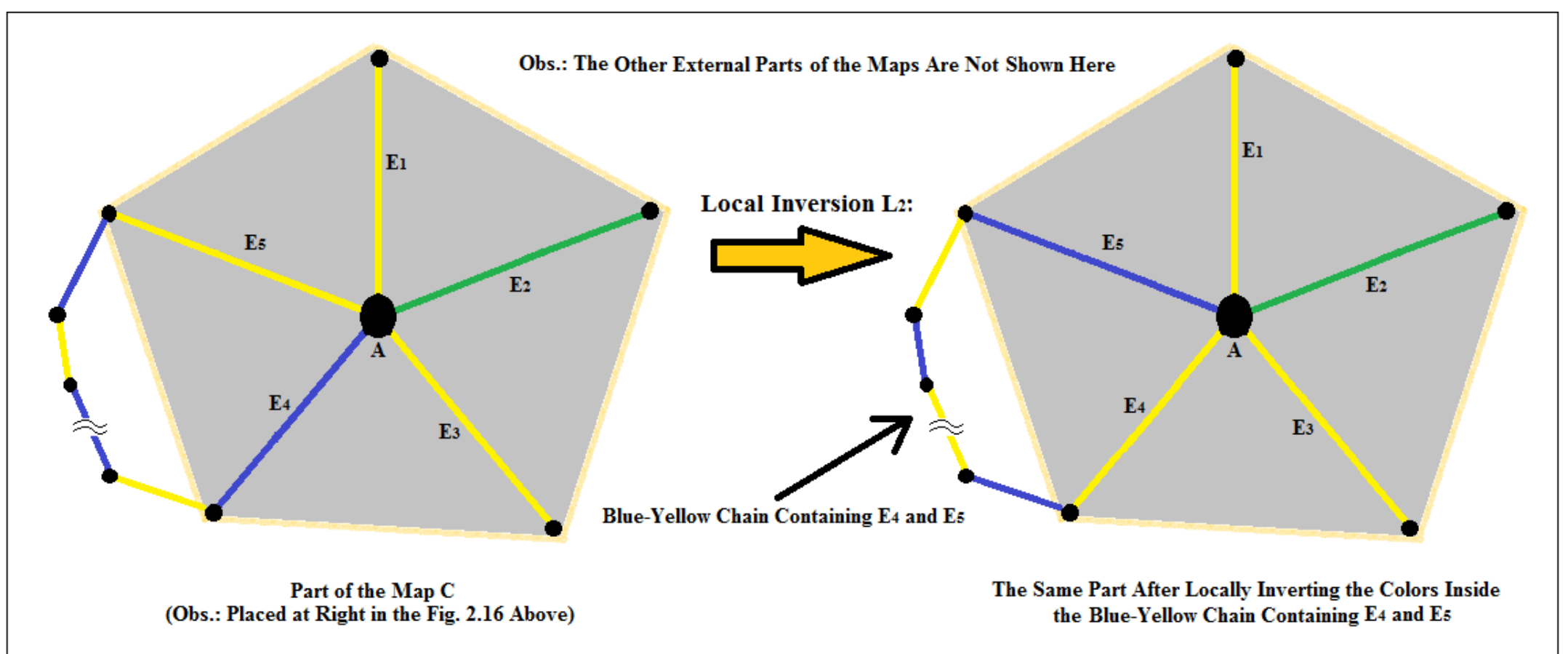

**Figure 2.23** Locally inverting the colors in a blue-yellow chain, then resulting a map that is <u>not</u> a *TBCI map*

Note that if the blue-yellow chain containing $E_5$ also contained $E_1$ (or $E_3$), then we could locally invert these colors in this chain $E_5$–$E_1$ (or $E_5$–$E_3$), generating a *TBCI map*, which also would imply that that map ***N*** could not be a minimal counter-example to the 4CT. Thus, in order to maintain our hypothesis, this blue-yellow chain (that contains $E_5$) must not contain either $E_1$ or $E_3$: From $E_1$, $E_3$ and $E_4$, it can contain only $E_4$. That is, that yellow-blue chain beginning in $E_5$ must end up in $E_4$ in the Fig. 2.23, since if it did it in $E_1$ or $E_3$, then a *TBCI map* would be generated when the colors from that yellow-blue chain (beginning in $E_5$ and ending up in either $E_1$ or $E_3$) were locally inverted (which also would imply that that map ***N*** could not be a true minimal counter-example to the 4CT).

So, as in the Fig. 2.10, the two possible fit topologies of that yellow curve from that resultant map ***C*** colored as at right in the Fig. 2.23 (where the edges $E_1$, $E_3$ and $E_4$ are yellow, $E_2$ is green, and $E_5$ is blue) are shown in the figure below:

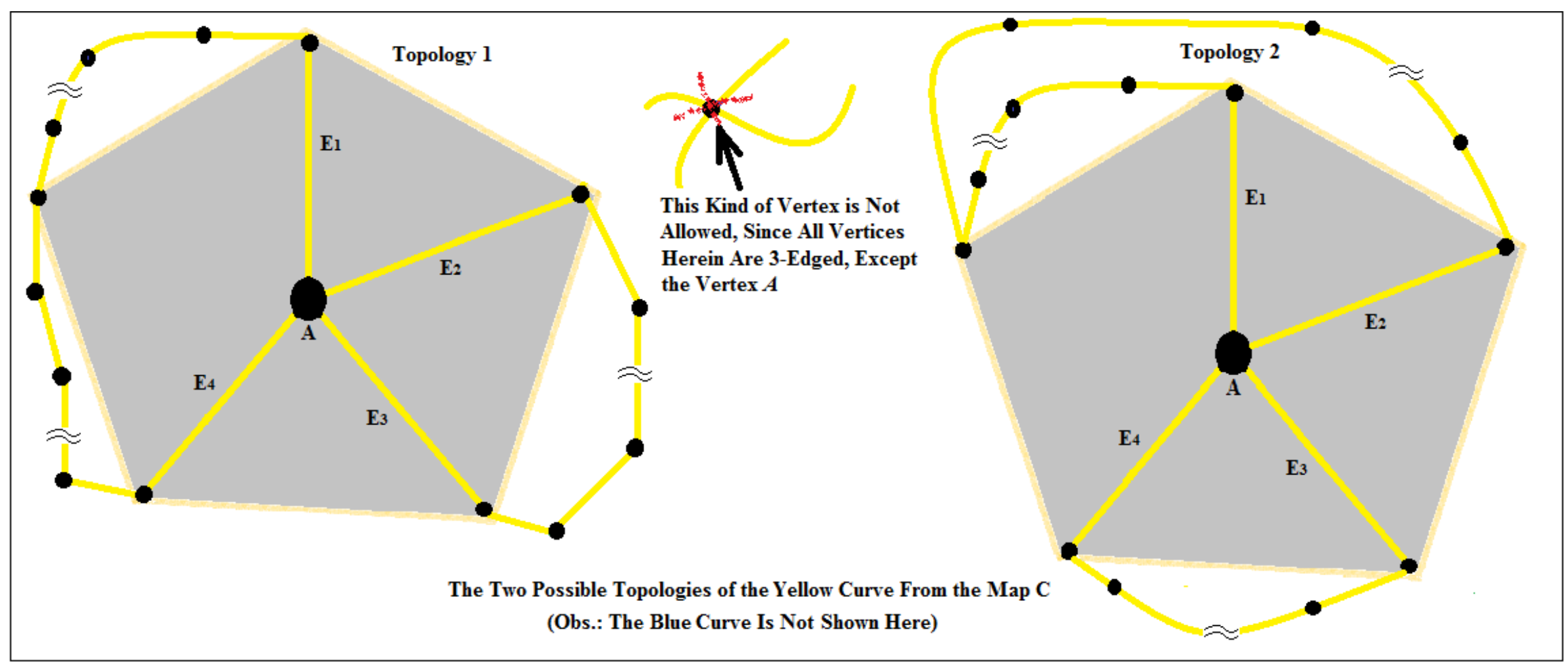

**Figure 2.24** Two possible topologies of a yellow curve of the map ***C*** that pass by vertex ***A***, after *inversion* $\boldsymbol{L_2}$

Again, as in the Fig. 2.11, if it was topology 2 (or 2'), then would exist a yellow-green chain including $E_1$ and $E_2$, which would allow a local inversion of these colors in this chain, generating a *TBCI map*, which also would imply that that map ***N*** could not be a real minimal counter-example to the 4CT, as shown in the figure below:

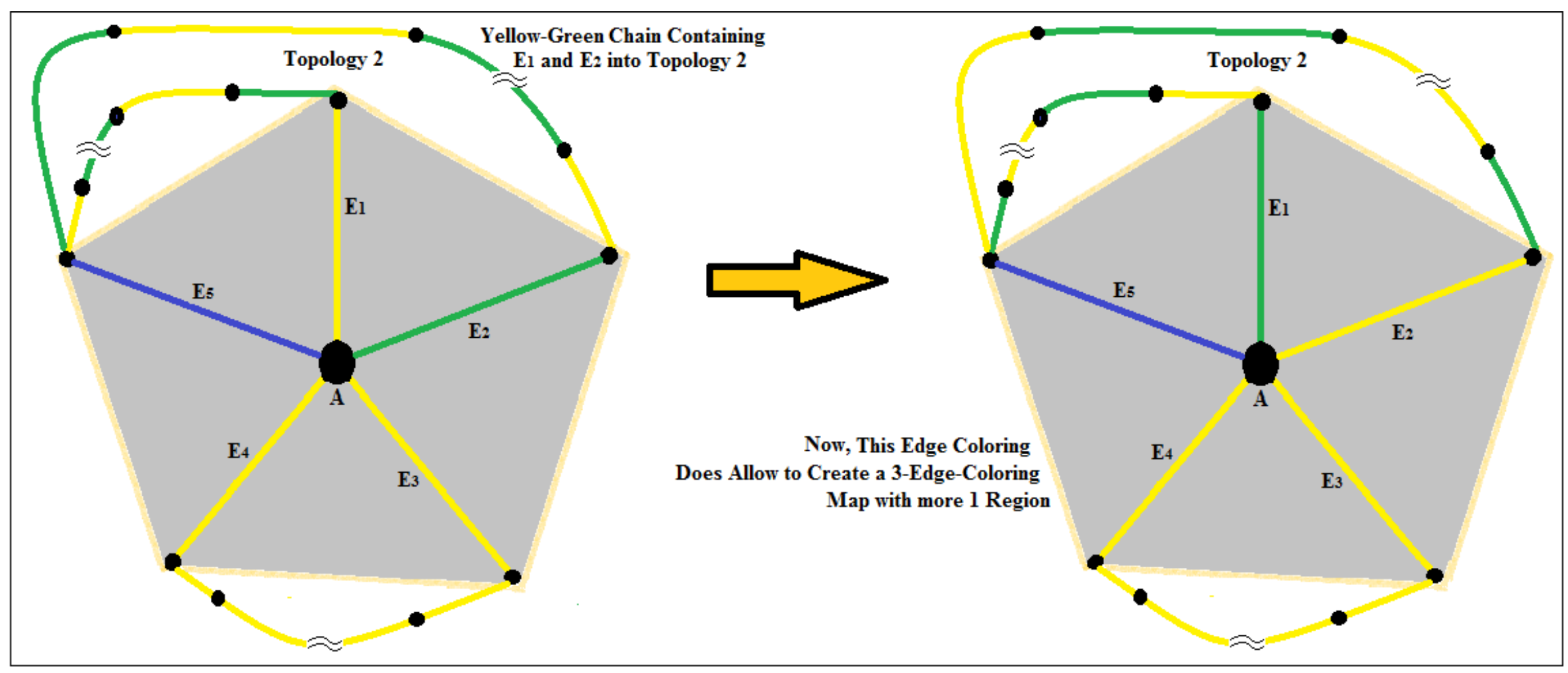

**Figure 2.25** Locally inverting colors in a yellow-green chain, then resulting a *TBCI map*

Hence, as that map ***N*** is a minimal counter-example to the 4CT, by hypothesis, that topology is (must be) the 1 (or 1'). Note that these two local inversions ($\boldsymbol{L_1}$ and $\boldsymbol{L_2}$) must be disjoint (they cannot have any edge in common, because a blue-yellow chain cannot cross another one in a 3-edged vertex, since there is only one pair of blue-yellow edges in each vertex of this type), which implies that that blue-yellow chain containing the edges $E_4$ and $E_5$, shown in the Fig. 2.23, cannot contain any edge in that blue-yellow chain containing the edges $E_1$ and $E_3$, shown in the Fig. 2.16.

Moreover, as shown in the Figs. 2.29 (I and III) (where $Y_{\{i,j\}}$ [$B_{\{i,j\}}$] denote the part of a yellow [blue] curve linking directly the vertices $V_i$ and $V_j$ without passing by ***A***), even though the blue-yellow chain containing the edges $E_4$ and $E_5$ crosses the green-yellow chain containing the edges $E_1$ and $E_2$ in the map, a local inversion $\boldsymbol{L_2}$ cannot cut the part of that yellow curve linking directly the edges $E_1$ and $E_2$ in that original map, in order to generate the topology 1 (or 1') again into another position in this map. This part of that yellow curve is just changed in order to pass by other vertices, but it continues linking directly the edges $E_1$ and $E_2$, as demonstrated in the Figs. 2.26 to 2.29 (III), where that case represents the general one.

However, in order to uphold our hypothesis (that map ***N*** is really a minimal counter-example to the 4CT), the local inversions $\boldsymbol{L_2}$ must cut the parts of that yellow curve linking directly the edges $E_1$ and $E_2$ (and $E_3$ and $E_4$) in that original map, in order to change the connections of the yellow curve *from* $E_1$ to $E_2$ *to* $E_2$ to $E_3$, and *from* $E_3$ to $E_4$ *to* $E_1$ to $E_5$, so as to generate the topology 1 (or 1') again into another position in the map, as it is required in our necessary sequence of topologies of the yellow curve after those local inversions $\boldsymbol{L_1}$ and $\boldsymbol{L_2}$, with the aim of upholding the truth of our hypothesis, as shown in the Figs. 2.26 to 2.29 (III):

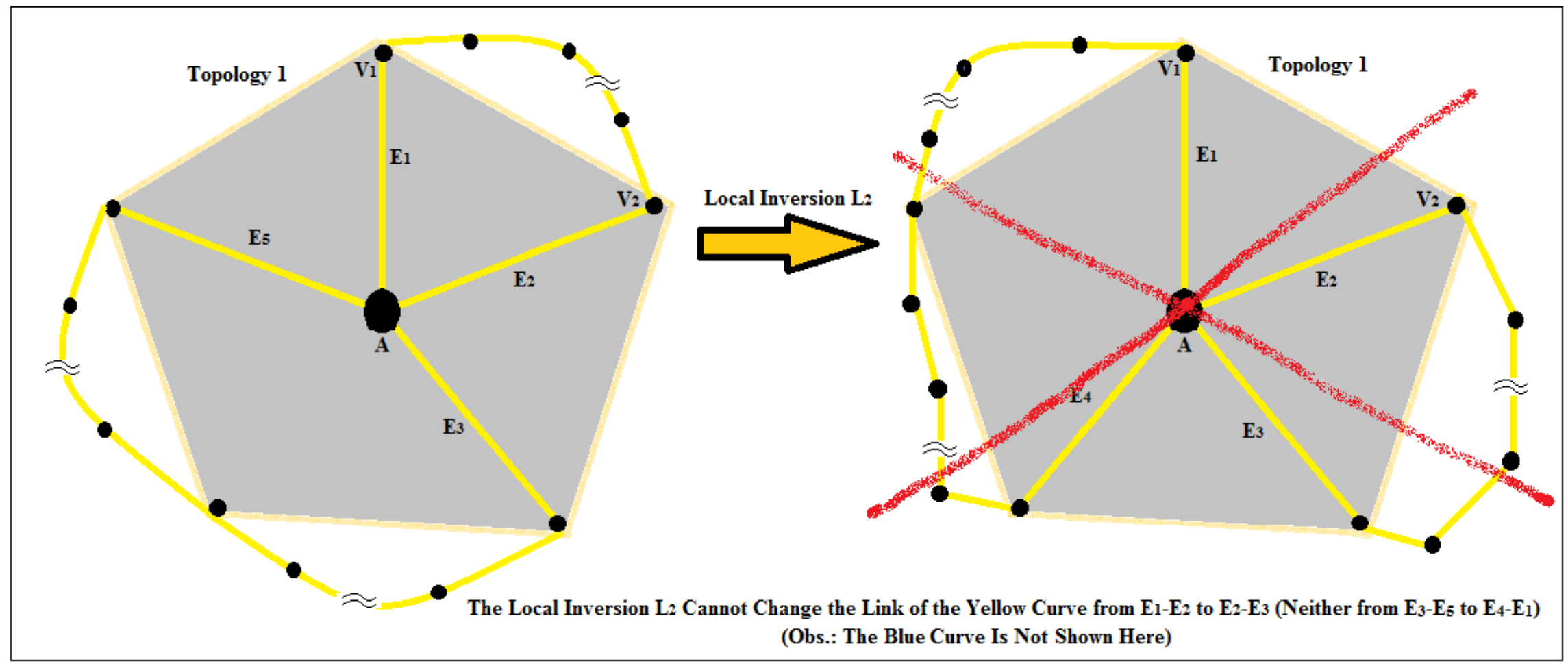

**Figure 2.26** The inversion $\boldsymbol{L_2}$ under a topology 1 (or 1') cannot generate this same kind of topology (1 or 1') where $V_1$ and $V_2$ are not directly linked by the yellow curve, as they are so in the left side above

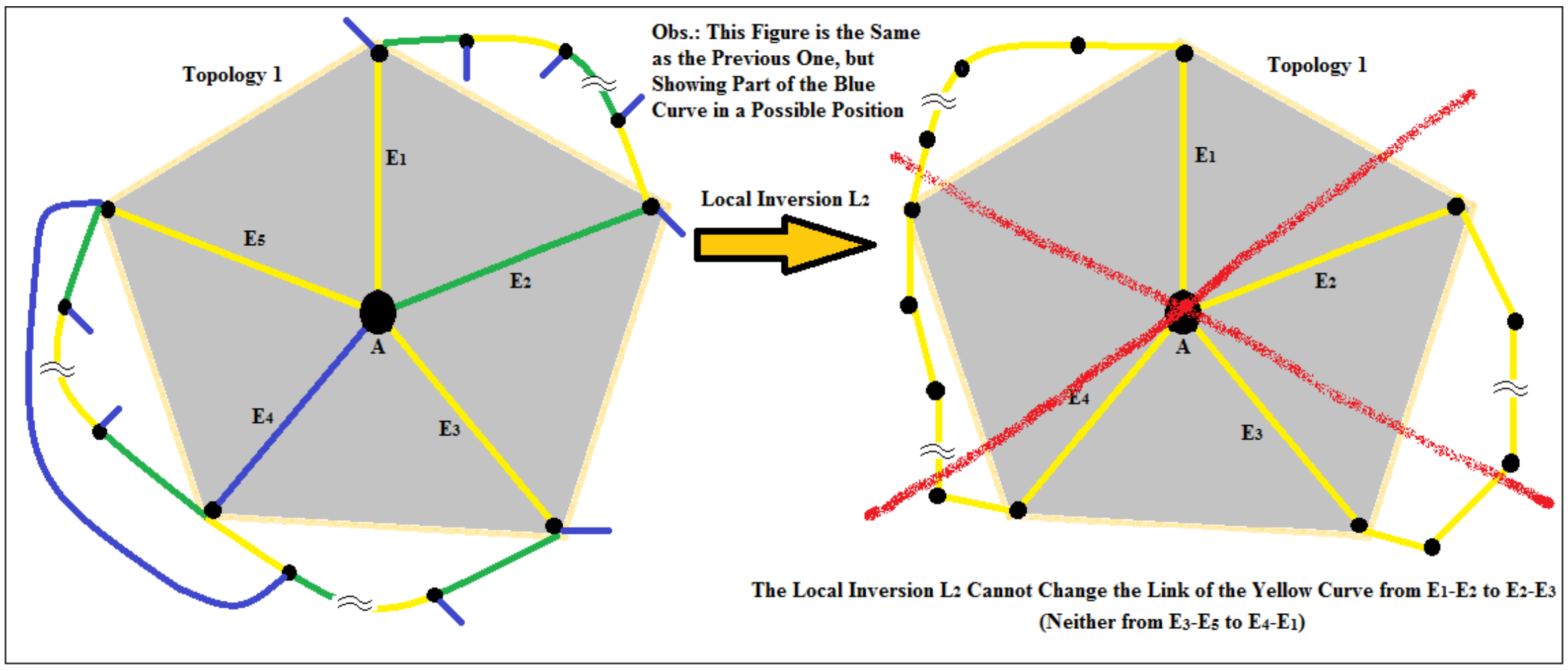

**Figure 2.27** The inversion $\boldsymbol{L_2}$ under a map with topology 1 cannot generate a map with topology 1 (or 1')

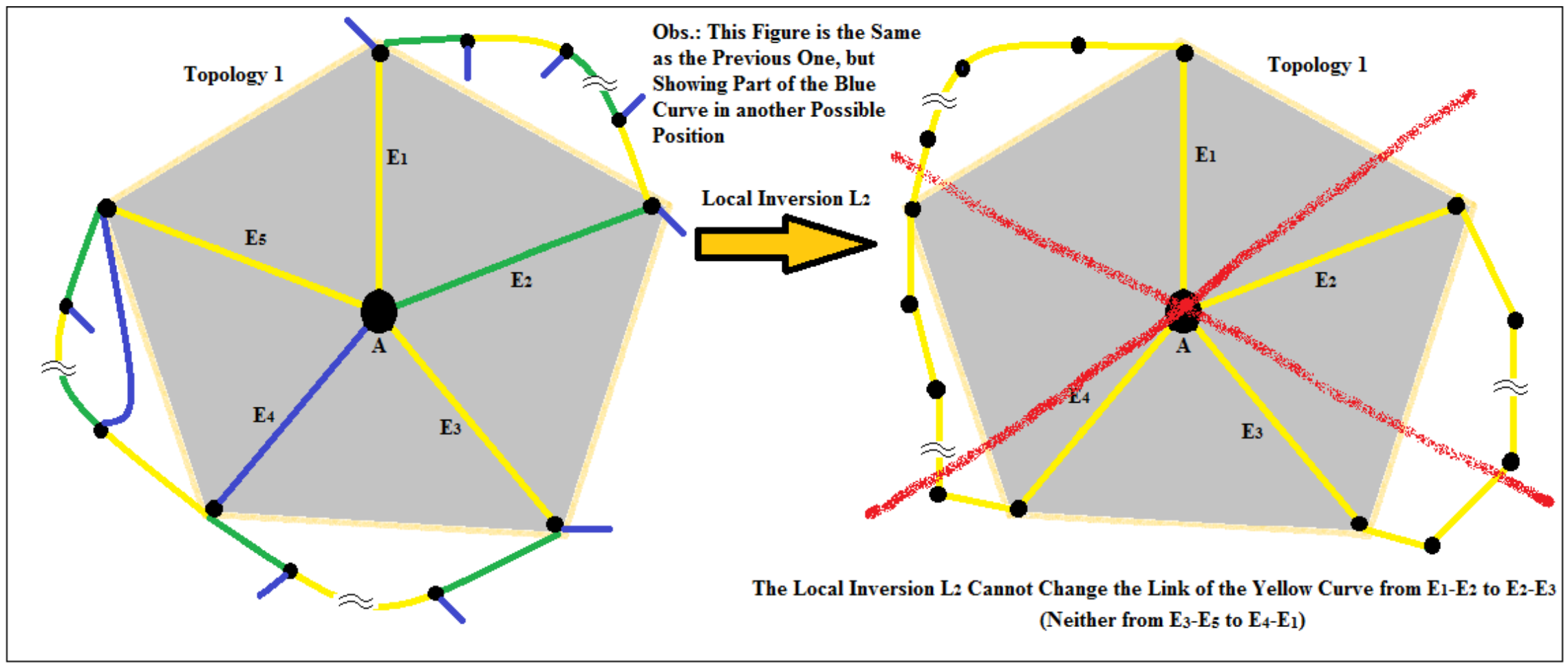

**Figure 2.28** The inversion $\boldsymbol{L_2}$ under a map with topology 1 cannot generate a map with topology 1 (or 1')

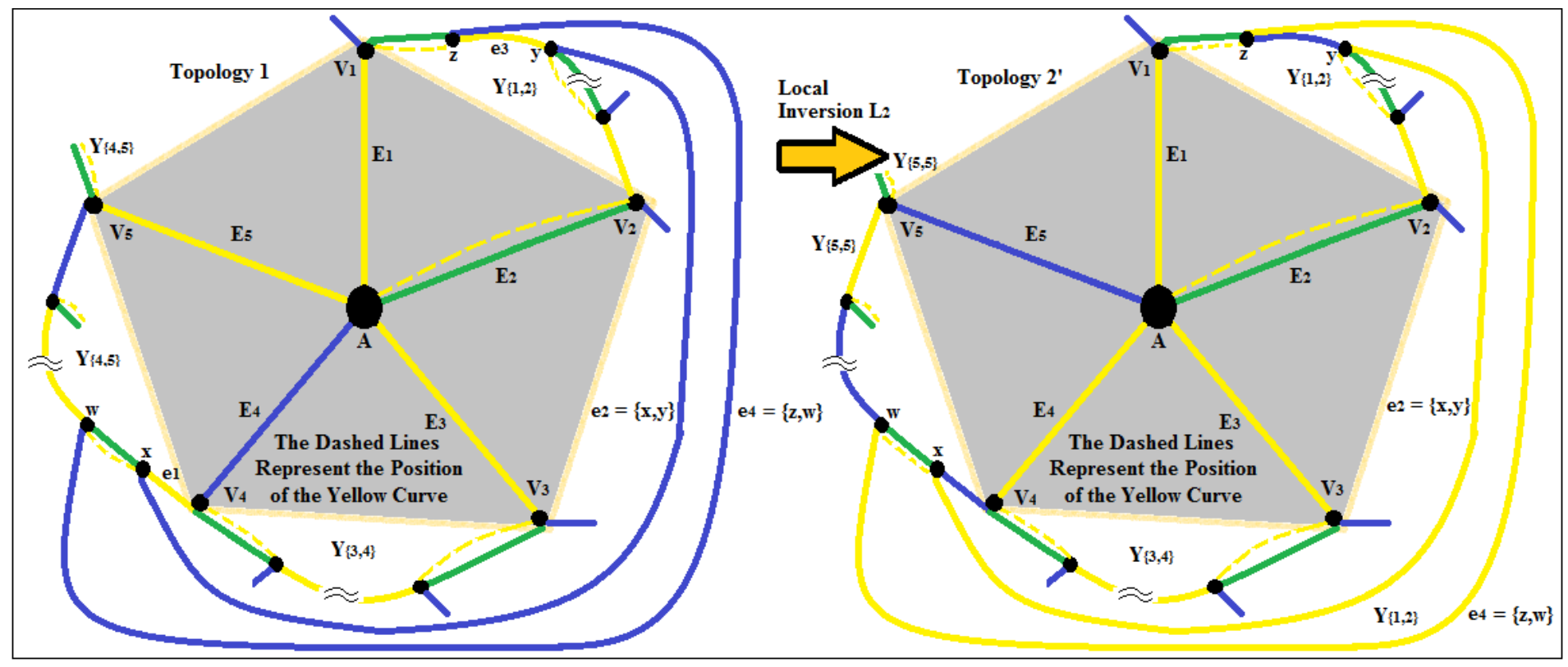

**Figure 2.29 (I)** The inversion $L_2$ under the previous map cannot generate a map with topology 1 (or 1')

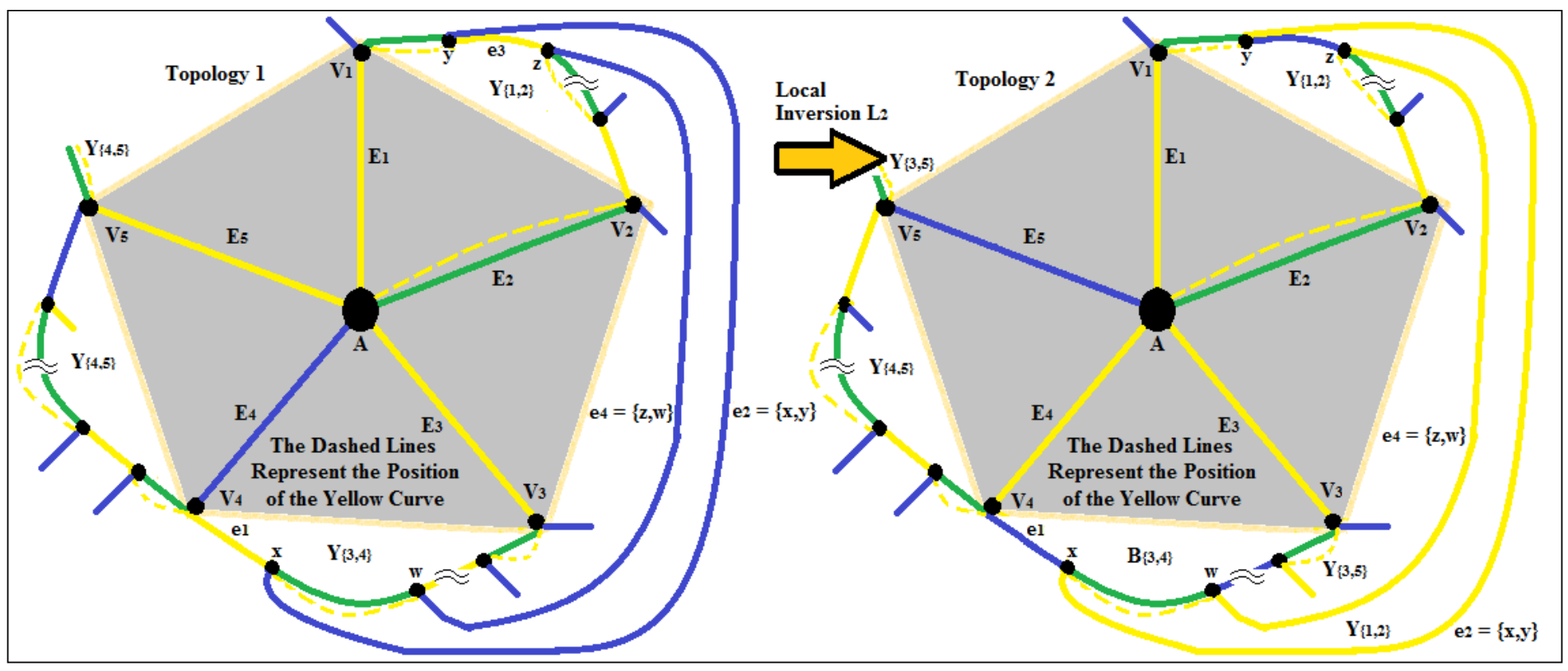

**Figure 2.29 (II)** The inversion $L_2$ under the previous map cannot generate a map with topology 1 (or 1')

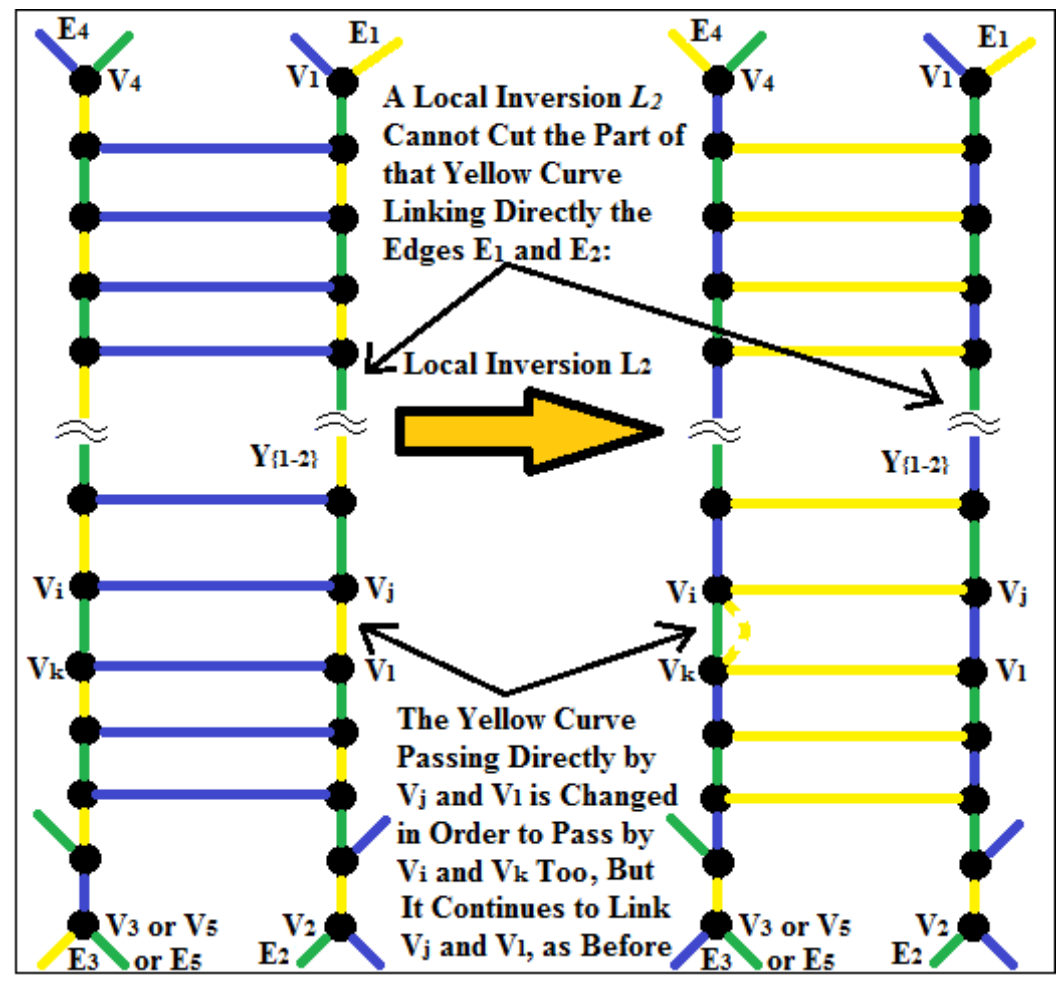

**Figure 2.29 (III)** The inversion $L_2$ under the previous map cannot generate a map with topology 1 (or 1')

As demonstrated above, even when that blue-yellow chain containing the edges $E_4$ and $E_5$ (or $E_3$) crosses the green-yellow chain containing the edges $E_1$ and $E_2$ in the map, the part of the yellow curve passing directly by arbitrary vertices $V_j$ and $V_l$ can be changed only in order to pass by $V_i$ and $V_k$ too (where $V_i$ was linked to $V_j$ and $V_k$ was linked to $V_l$ by a blue curve), but those vertices $V_j$ and $V_l$ continue linked by that yellow curve. As $V_j$ and $V_l$ are arbitrary vertices and form a direct link from $E_1$ to $E_2$, and that local inversion $L_2$ can only invert the colors from the blue-yellow chain linking $E_4$ to $E_5$ (or $E_3$), then this local inversion cannot cut the part of that yellow curve linking directly the edges $E_1$ and $E_2$.

Therefore, our hypothesis (that map ***N*** is a minimal counter-example) is (must be) false, and a minimal counter-example ***N*** do not exist, since all the allowed colorings in the Table 2.4 lead to the fact that that map ***N*** is (must be) a *3-ECC* (so, also a *4-CM*).

So, in synthesis, if the local inversion $\boldsymbol{L_1}$ under that *CBG* ***C*** (originated from that arbitrary map ***N***) generates a map with topology 2 (or 2'), then we can invert the colors in some yellow-green chain (by the Lemma 2.7, since in this case, e.g. the edges $E_2$ and $E_3$ are kept in a simple cycle forming a yellow-green chain), generating a *TBCI map*, which would imply that that map ***N*** could not be a real minimal counter-example to the 4CT, denying our hypothesis, as shown in the figure below:

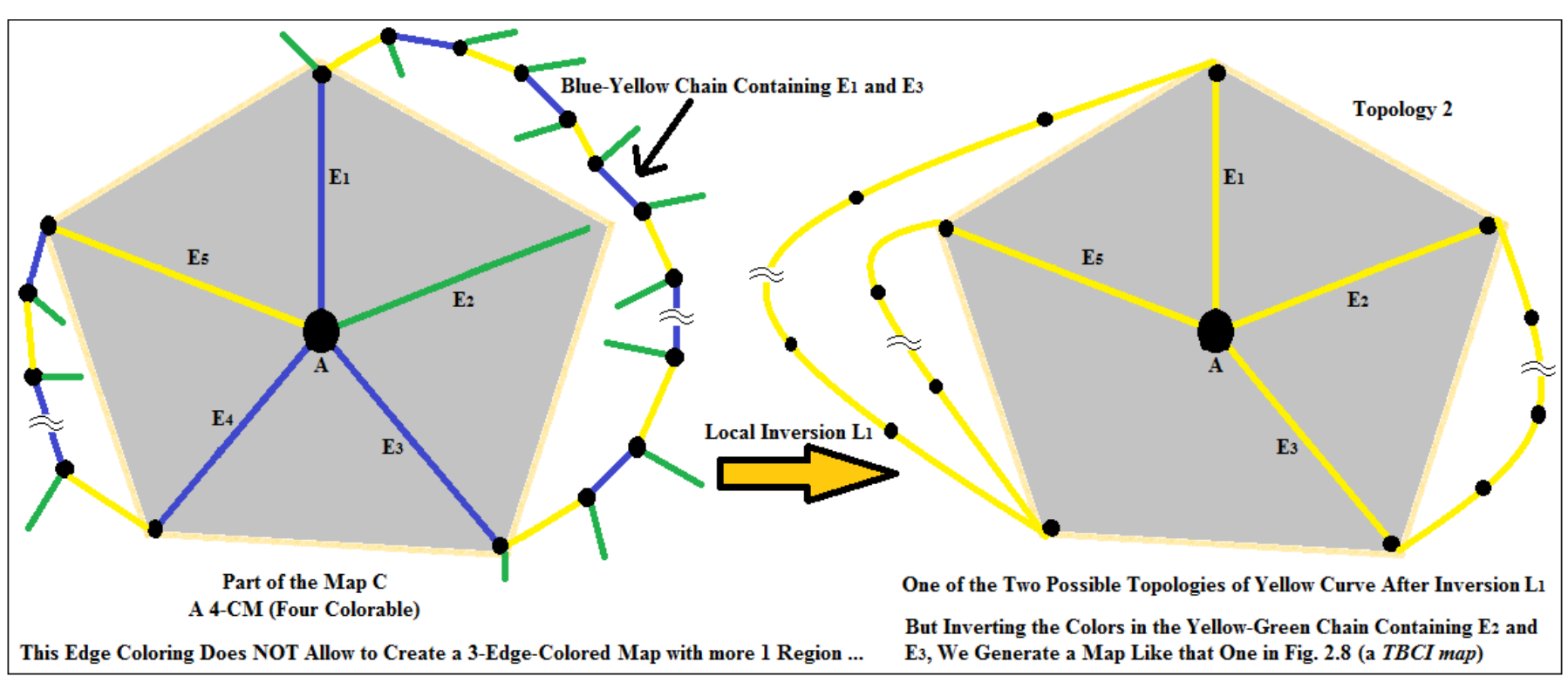

**Figure 2.30** If local inversion $\boldsymbol{L_1}$ leads to topology 2 (or 2'), then ***N*** cannot be a minimal counter-example

Hence, as the local inversion $\boldsymbol{L_1}$ under that map ***C*** cannot generate a map with yellow curve with topology 2 (or 2'), in order to maintain the truth of our hypothesis, it can only generate a map with yellow curve with topology 1 (or 1'), as shown in the figure below:

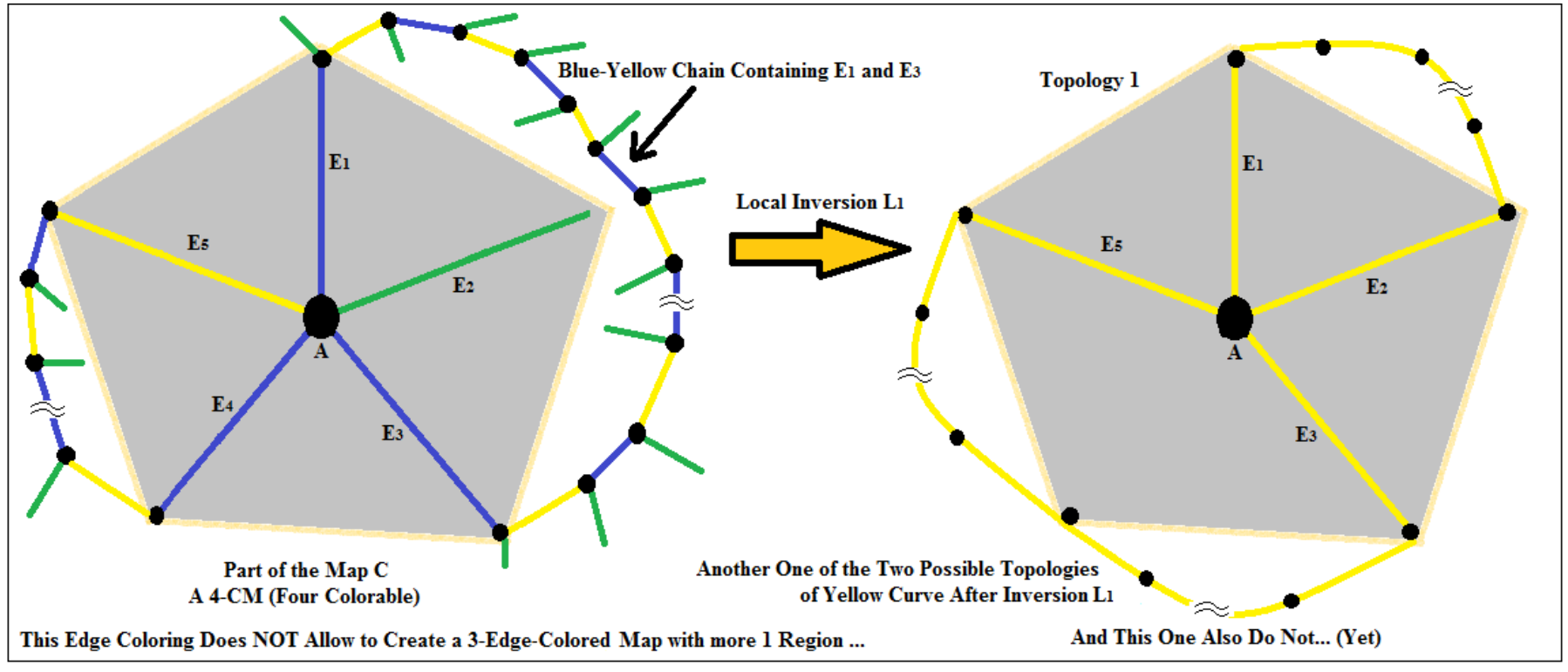

**Figure 2.31** If inversion $\boldsymbol{L_1}$ leads to topology 1, then the map ***N*** can be a minimal counter-example (yet)

Now, after the inversion $\boldsymbol{L_1}$ under that map ***C***, we must have certainly a map with yellow curve with topology 1 (or 1'). Then, after a local inversion $\boldsymbol{L_2}$ under that map ***C*** we have certainly a map with yellow curve with topology 2 (or 2'), since – as we have seen above – the inversion $\boldsymbol{L_2}$ under that map ***C*** cannot from a map with yellow curve with topology 1 (or 1') create a map with yellow curve with this same topology 1 (or 1'), into another position:

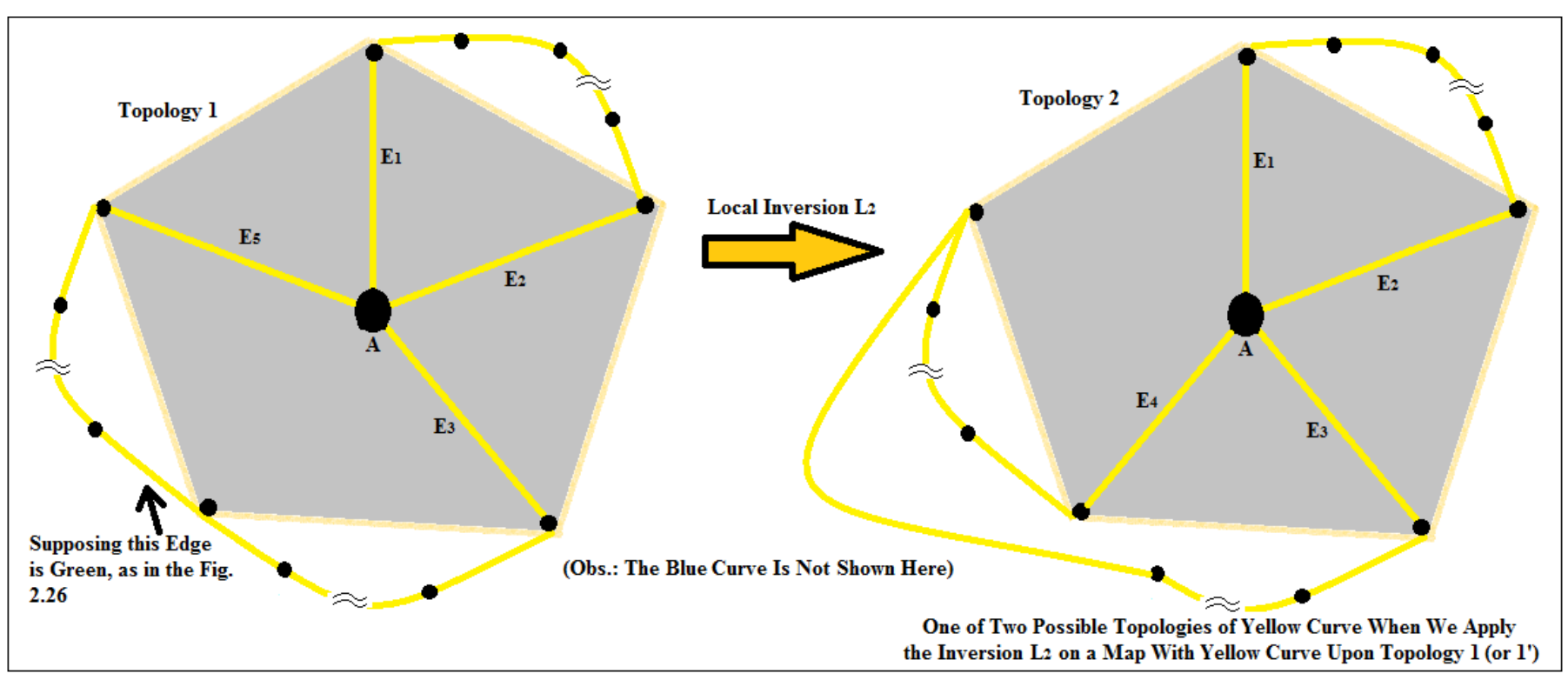

**Figure 2.32** The inversion $\boldsymbol{L_2}$ under a map with topology 1 can only generate a map with topology 2 (or 2')

Then, from that generated map with topology 2, as already shown above, we can generate a *TBCI map*, doing a local inversion $L_2$, where that map $N$ is proved cannot be an actual minimal counter-example to the 4CT, as already shown previously.

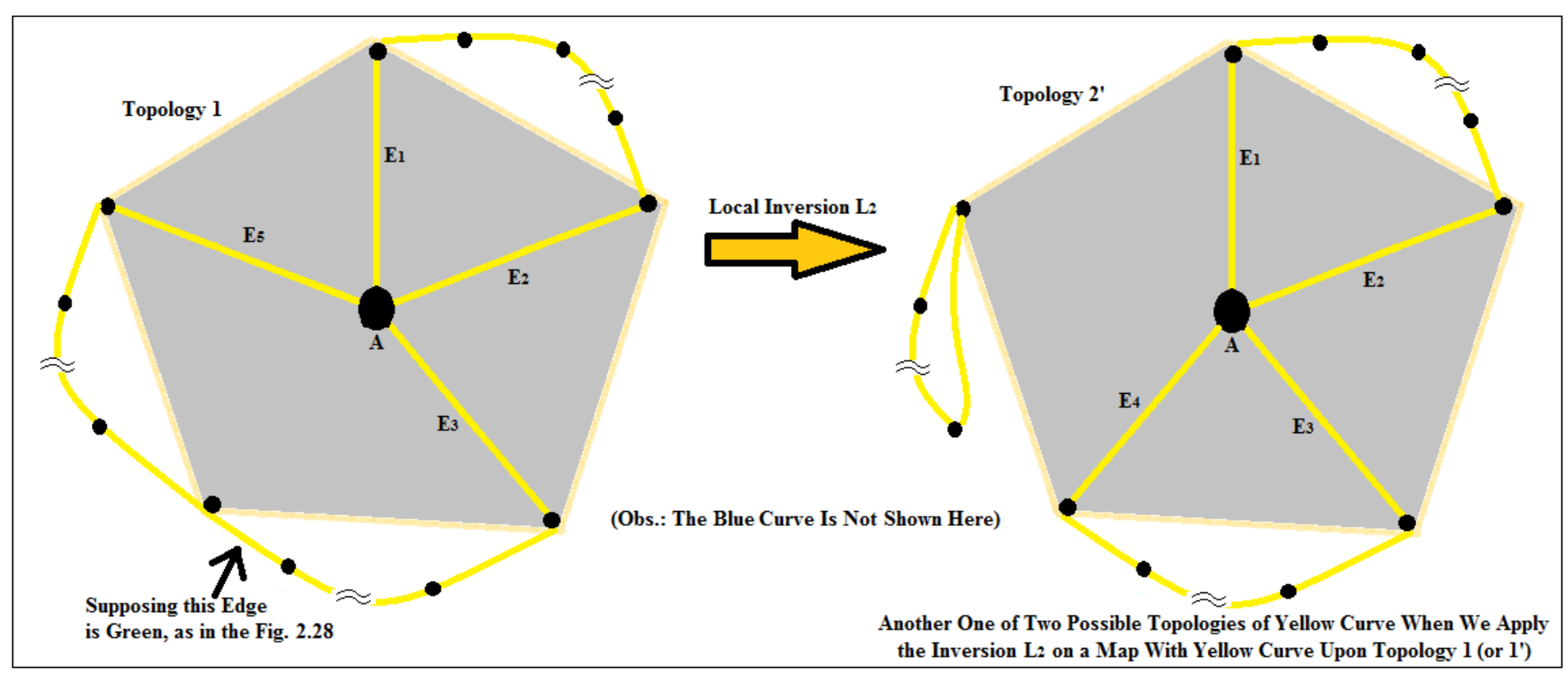

**Figure 2.33** The inversion $L_2$ under a map with topology 1 can only generate a map with topology 2' (or 2)

Finally, from that generated map with topology 2', as already shown above w.r.t. topology 2, we can also generate a *TBCI map*, executing also a local inversion $L_2$, where is again proved that that map $N$ cannot be a true minimal counter-example to the 4CT.

Proof brief: A [supposed] arbitrary minimal counter-example *CBG* $N$ => Contracting pentagon => A *4-CM* map $C$ that is <u>not</u> a *TBCI map* => Topology 1 (or 1') => Local inverting $L_1$ => Topology 1 (or 1') => Local inverting $L_2$ => Topology 2 (or 2') => A *4-CM* map $C$ that <u>is</u> a *TBCI map* => $N$ is <u>not</u> a minimal counter-example => 4CT (by a short and simple proof that can be utterly checkable by human mathematicians, without computer assistance).

Hence, a true minimal counter-example to the 4CT does not exist, hence every connected finite simple planar graph is a ***3-ECC***, so it is also a ***4-CM***, and hence **4CT** is proved. ☐

## 3. Conclusion & Understanding

The conclusion is that now we utterly understand [12] why the 4CT is really true: Every finite simple planar graph can be properly four-colorable because all they can be represented by only two sets of closed curves, where all the regions of that map is either inside or outside with respect to each one of these two sets of curves, and when we cross any edge (entering at some adjacent region) at least one of these relative positions must change.

So, a properly four-coloring emerges naturally, in every finite map, associating every region of the map to each one of 4 arbitrary colors biunivocally associated to the 4 possible relative positioning of that region with respect to two set of closed curves that form that map:

| Color of the Region | Location w.r.t. *DSCC* 1 | Location w.r.t. *DSCC* 2 |
|---|---|---|
| **1** | Outside | Outside |
| **2** | Outside | Inside |
| **3** | Inside | Outside |
| **4** | Inside | Inside |

**Table 3.1** Understanding utterly why the 4CT is really true

## 4. Comments about Interesting Reviews

4.1 A reviewer has said:

"On page 18, in the last paragraph, the author states "... the local inversion $L_2$ cannot cut the part of that yellow curve linking directly the edges $E_1$ and $E_2$ in that original map, in order to generate the topology 1 (or 1') again into another position in this map..." But apparently exactly this might happen! Let me illustrate such a situation: – We assume that map C has edges $E_1$, $E_3$, and $E_5$ in yellow, edge $E_2$ in yellow and blue (hence green), and edge $E_4$ in blue, as illustrated in the right of Figure 2.16.

– Moreover, we assume that map C has Topology 1 as illustrated in the left of Figure 2.20.

– For future reference, for i=1,...,5, let $v_i$ denote the endpoint of edge $E_i$ that is different from A.

– Let $Y_{\{1,2\}}$ denote the yellow path between $v_1$ and $v_2$ (see left of Fig. 2.20). Similarly, let $Y_{\{3,4\}}$ denote the yellow path between $v_3$ and $v_4$ (also in the left of Fig. 2.20). Yet similarly, let $Y_{\{4,5\}}$ denote the yellow path between $v_4$ and $v_5$ (again in the left of Fig. 2.20).

– Now consider the path P used in the local inversion $L_2$. That is, P has endpoints $v_4$ and $v_5$, does not contain vertex A, and alternates in colors yellow and blue. The author claims that swapping colors yellow and blue on this path cannot result in a Topology 1 situation, which is depicted in the left of Figure 2.24.

– However, exactly this would happen if path P is routed as follows: Starting from $v_4$, take one (yellow) edge $e_1 = \{v_4,x\}$ of $Y_{\{3,4\}}$, take one blue edge $e_2 = \{x,y\}$ whose other endpoint y is on $Y_{\{1,2\}}$, take one (yellow) edge $e_3 = \{y,z\}$ of $Y_{\{1,2\}}$ in direction towards $v_1$, and take one last blue edge $e_4 = \{z,v_5\}$ whose other endpoint is $v_5$.

– Swapping yellow and blue colors along P and also on $E_4$ and $E_5$ (i.e., performing local inversion $L_2$), results in a yellow curve from vertex A along $E_1$, then along $Y_{\{1,2\}}$ up to vertex z, then along $e_4$ (which is now yellow) to vertex $v_5$, then along $Y_{\{4,5\}}$ up to vertex $v_4$, and finally along $E_4$ (which is now yellow) back to vertex A. (Consequently, there is a yellow curve A->$E_2$->$Y_{\{1,2\}}$->$e_2$->$Y_{\{3,4\}}$->$E_3$.) Hence, after local inversion $L_2$ we are facing Topology 1 again."

In the implicit topologies of the curves that the reviewer has utilized above, the blue edge $\mathbf{E_{b1}}$ from the vertex **w** in Fig. 2.34 (where it is represented which he/she has said) would be blocked by those edges $\mathbf{e_1}$, $\mathbf{e_2}$, $\mathbf{e_3}$ and $\mathbf{e_4}$, impeding that it could reach to the blue edge $\mathbf{E_{b2}}$ from $\mathbf{V_2}$, hence impeding that the closed blue curve that pass by the edges $\mathbf{E_2}$ and $\mathbf{E_4}$ could really be closed one.

Then, the referred local inversion used in that review cannot occur in a real map, where the supposed contradiction with the proof herein does not exist (by the way, see that in the local inversions that can really occur, that closed blue curve that pass by the edges $\mathbf{E_2}$ and $\mathbf{E_4}$ can (must) really be closed one, as seen in following Fig. 2.35). See yet, in Figs. 2.36 and 2.37, that even in a general way that construction from that review doesn't work:

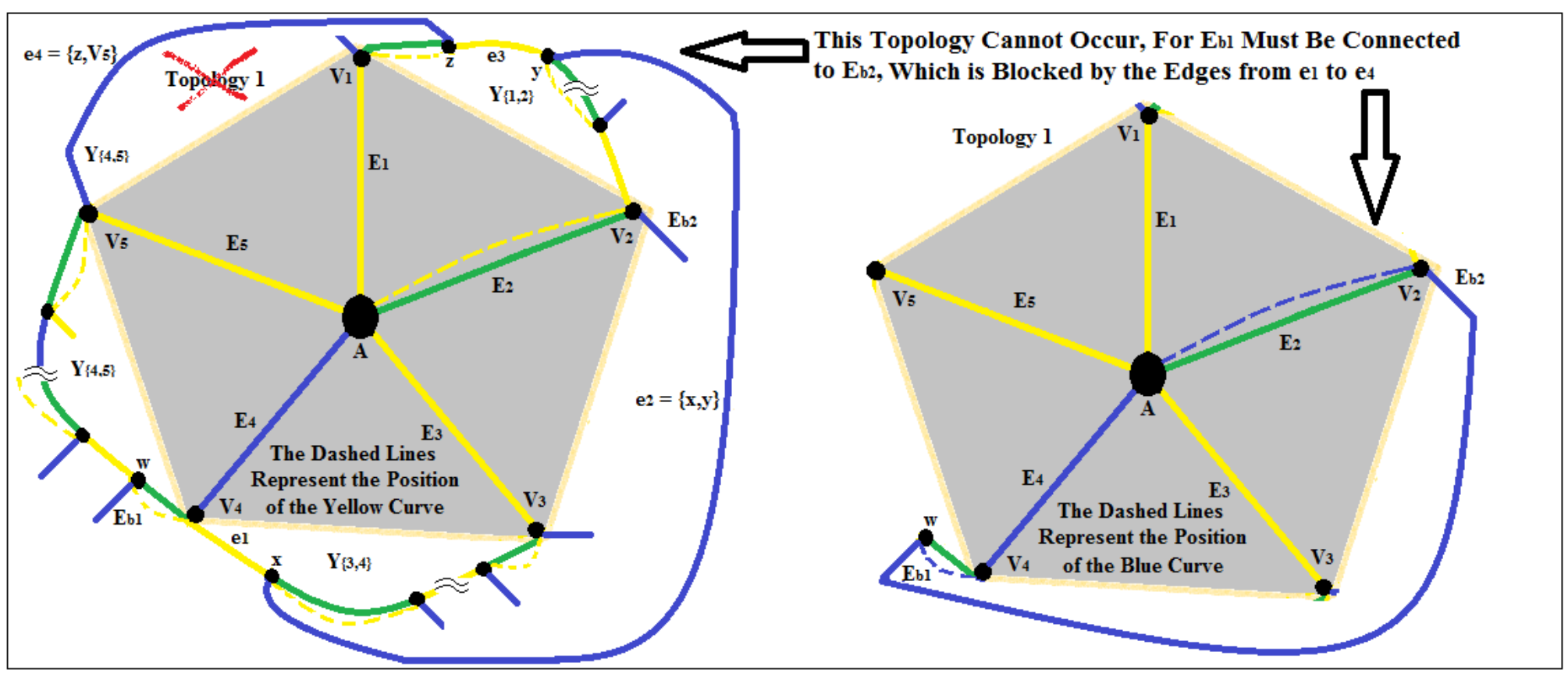

**Figure 2.34** Representation of the inconsistent Topology 1 used by the reviewer that has caused the error in that local inversion, as explained above

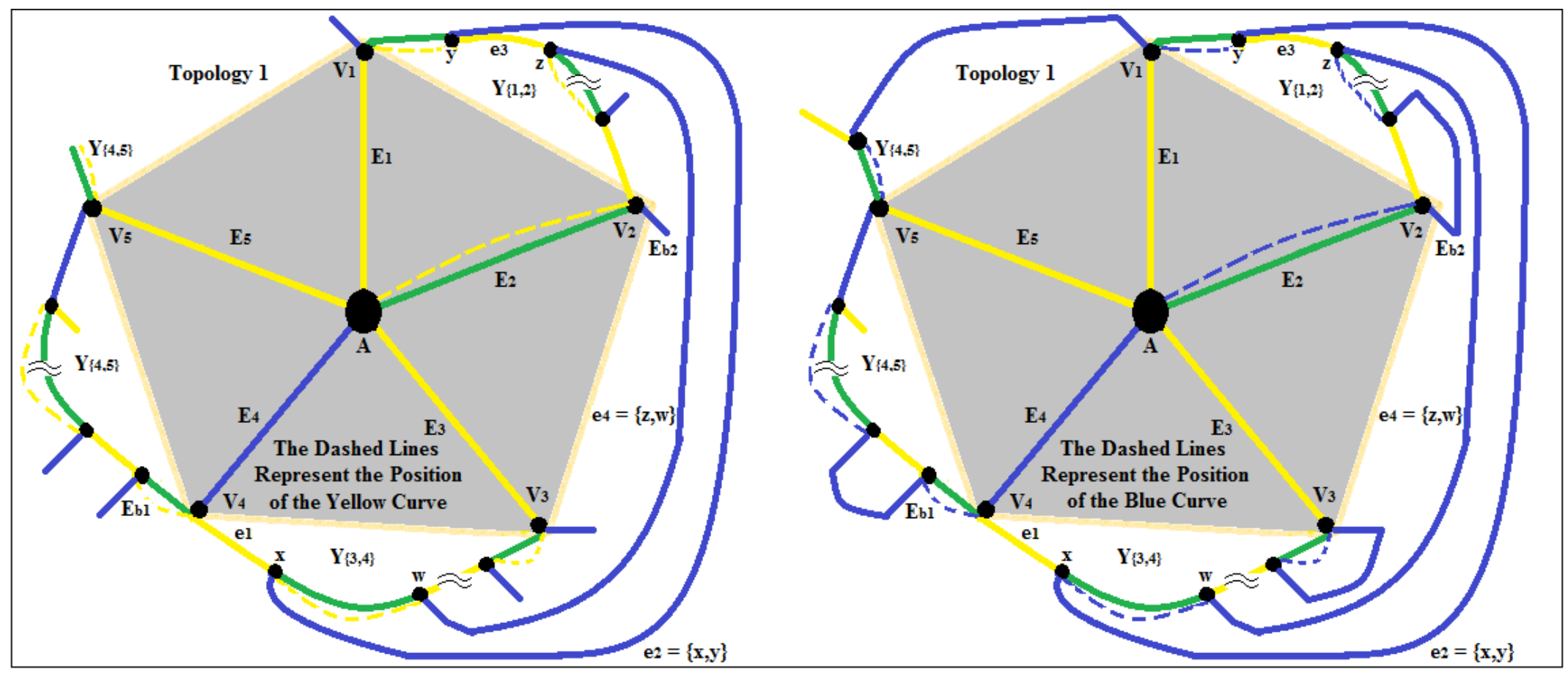

**Figure 2.35** Representation of the consistent Topology 1 used by the author [in the Fig. 2.29 (II)] in order to do a right local inversion

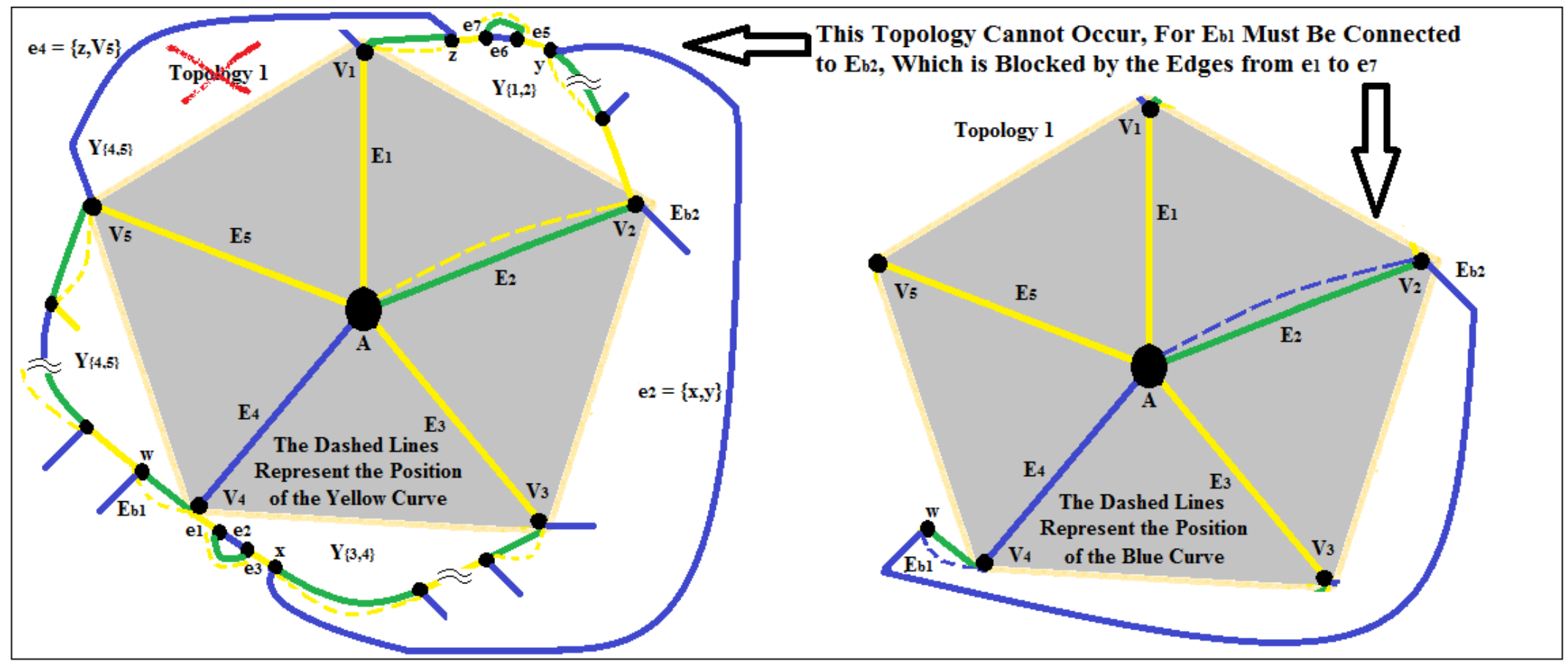

**Figure 2.36** Representation of a more general inconsistent Topology 1 causing error in that local inversion too

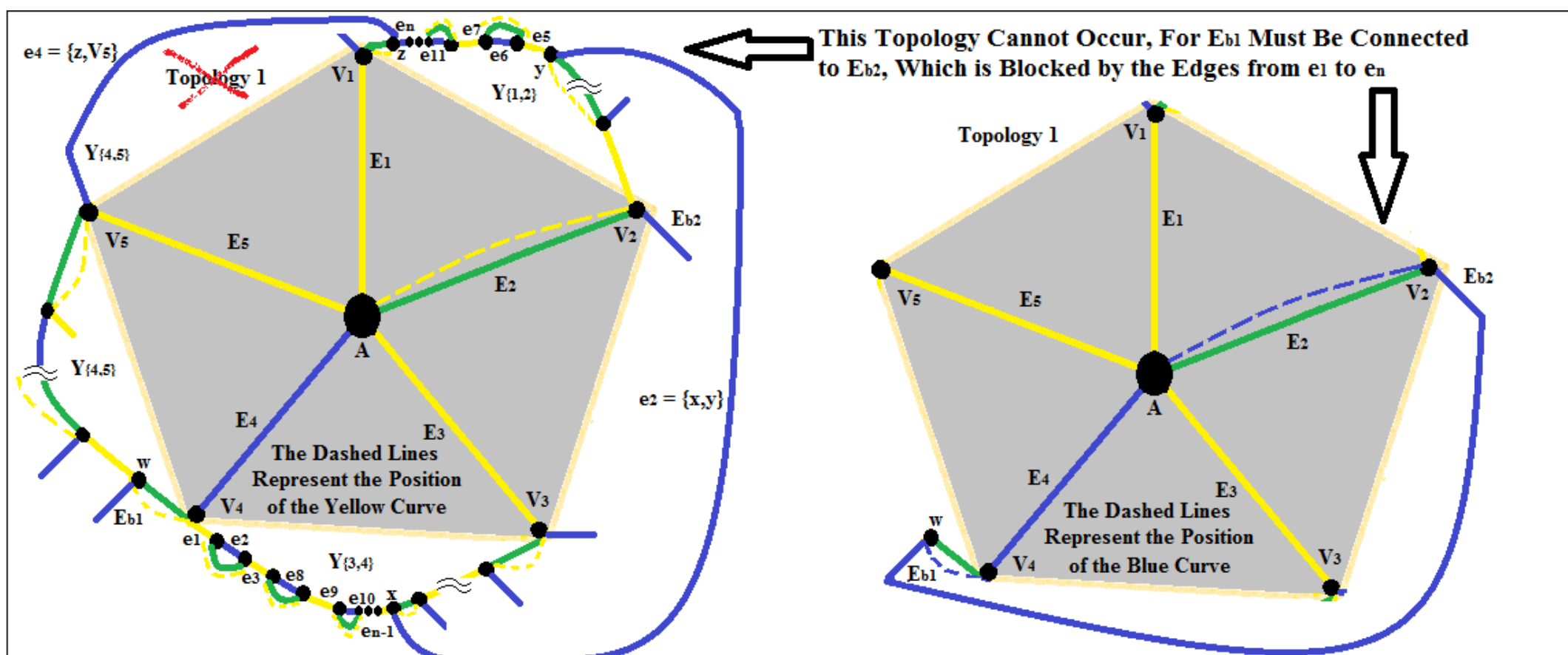

**Figure 2.37** Representation of a general inconsistent Topology 1 causing error in that local inversion too, definitely demonstrating that from a topology 1 (or 1'), after a local inversion $L_2$, we cannot obtain again a topology 1 (or 1') into another position in the map

## 4.2 Another reviewer has said:

"The author claims to have given a human comprehensible proof of the four colour theorem. Since the theorem itself is correct, as has been shown by the Appel & Haken computer generated proof and its subsequent simplifications, it cannot be excluded that such a proof might exist." Then, after to agree in general with all the other proof arguments, that reviewer nevertheless concludes: "However, I see no convincing argument that the blue-yellow chain through $E_4$ and $E_5$ in Fig. 2.23 must return to $E_5$; it may reenter in node ***A*** passing by $E_3$ or $E_1$."

See the cited Fig. 2.23 copied and pasted below, as Fig. 2.38:

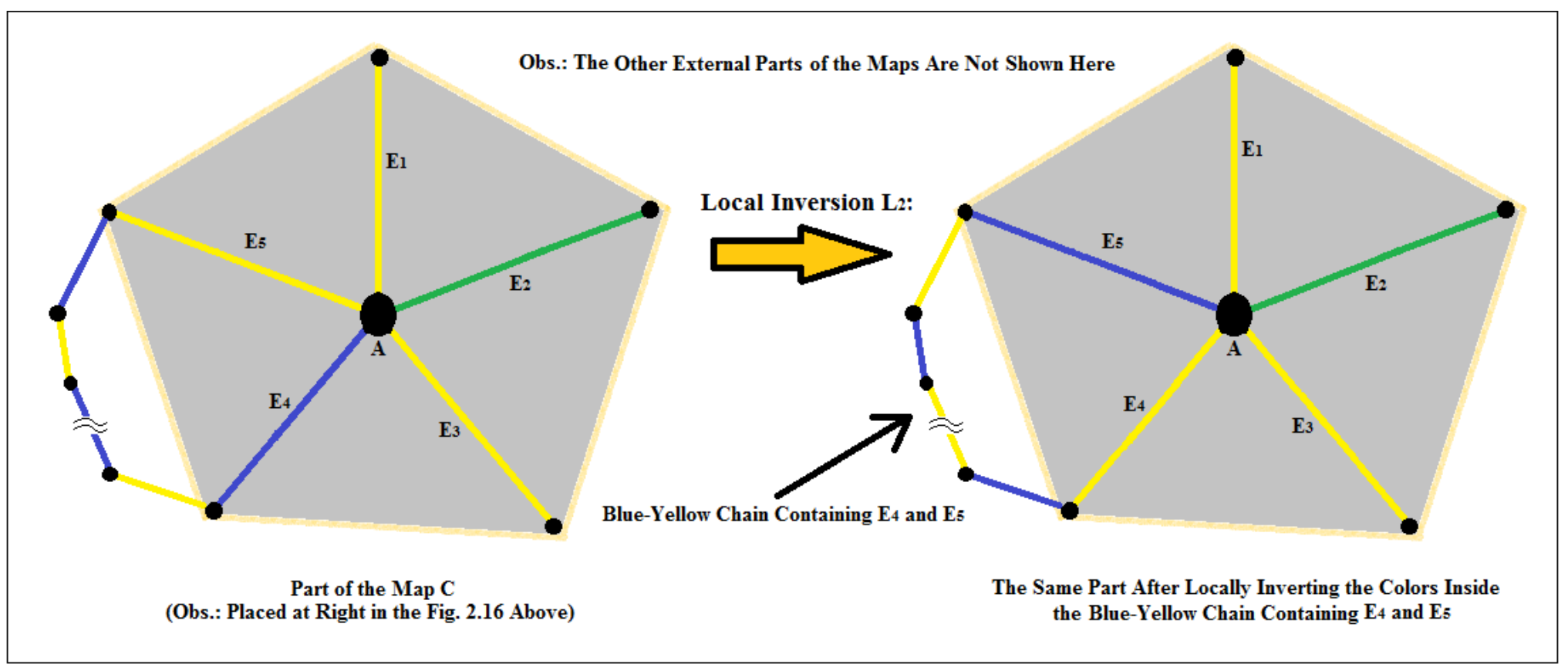

**Figure 2.38** Locally inverting the colors in a blue-yellow chain, then resulting a map that is not a *TBCI map*

So, we shall see that reviewer is not completely right: As demonstrated in Fig. 2.15, the blue-yellow chain through $E_4$ cannot reenter in node ***A*** passing by $E_1$, for if so, then it would block the blue-yellow chain containing $E_3$ and $E_5$, by planarity, as demonstrated in Fig. 2.15; and if it reenters in ***A*** passing by $E_3$, then there would be a local inversion of colors upon this blue-yellow chain that would generate a *TBCI map*, as in Fig. 2.14, which also would demonstrate that that map ***N*** could not be a true minimal counter-example to the 4CT:

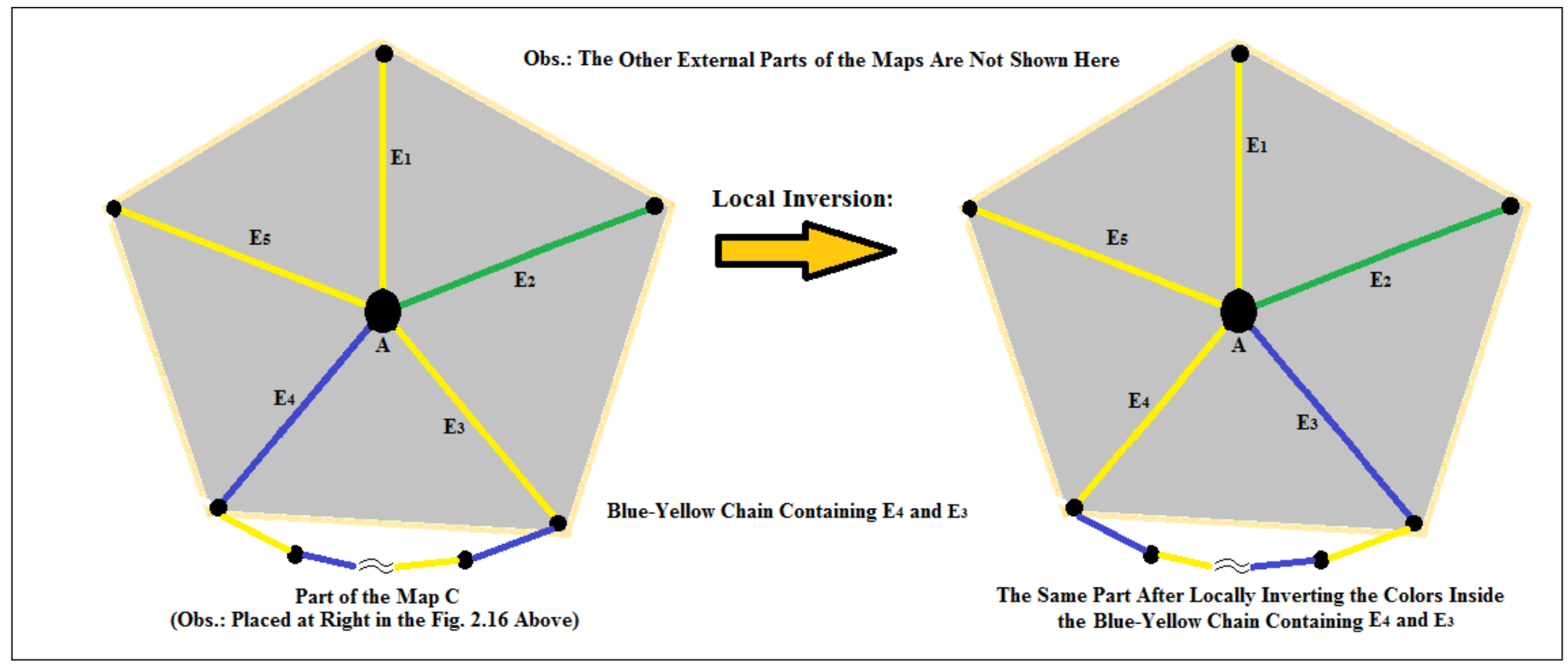

**Figure 2.39** Locally inverting the colors in a blue-yellow chain, then resulting a map that is a *TBCI map*

## 5. Freedom & Mathematics

"**– The essence of Mathematics is Freedom.**" (Georg Cantor) [11]

**André Luiz Barbosa – Goiânia - GO, Brazil – e-Mail:** webmaster@andrebarbosa.eti.br **– April 2016**

Site......... : www.andrebarbosa.eti.br
Blog......... : blog.andrebarbosa.eti.br

This Paper : http://www.andrebarbosa.eti.br/A_Human_Checkable_Four_Color_Theorem_Proof.htm
PDF……. : http://www.andrebarbosa.eti.br/A_Human_Checkable_Four_Color_Theorem_Proof.pdf
arXiv…… : https://arxiv.org/ftp/arxiv/papers/1708/1708.07442.pdf